\documentclass[twocolumn]{aastex631}
\usepackage{hyperref}
\usepackage{changepage}
\hypersetup{
    colorlinks=true,
    linkcolor=blue,
    filecolor=magenta,      
    urlcolor=cyan,
}

\usepackage{xspace}
\usepackage{float}
\usepackage{ucs}
\usepackage[utf8x]{inputenc}
\usepackage[version=4]{mhchem}
\usepackage{csvsimple}

\usepackage[caption=false]{subfig}
\usepackage[T1]{fontenc}

\newcommand{\kms}{\ensuremath{\mathrm{km}\,\mathrm{s}^{-1}}\xspace}

\newcommand{\gcc}{g~cm$^{-3}$\xspace}
\newcommand{\Rsun}{\ensuremath{R_{\odot}}\xspace }
\newcommand{\Msun}{\ensuremath{M_{\odot}}\xspace}
\newcommand{\Lsun}{\ensuremath{L_{\odot}}\xspace}

\newcommand{\tess}{\textit{TESS} }
\newcommand{\kepler}{\textit{Kepler} }
\newcommand{\jwst}{\textit{JWST} }
\newcommand{\MEarth}{\ensuremath{M_{\oplus}}\xspace}
\newcommand{\REarth}{\ensuremath{R_{\oplus}}\xspace}
\newcommand{\lo}{$\ell_1$\xspace}

\begin{document}

\nolinenumbers
\title{Early Results from the HUMDRUM Survey: A Small, Earth-mass Planet Orbits TOI-1450A
}

\correspondingauthor{Madison Brady}
\email{mtbrady@uchicago.edu}

\author[0000-0003-2404-2427]{Madison Brady}
\affiliation{Department of Astronomy \& Astrophysics, University of Chicago, Chicago, IL 60637, USA}

\author[0000-0003-4733-6532]{Jacob L.\ Bean}
\affiliation{Department of Astronomy \& Astrophysics, University of Chicago, Chicago, IL 60637, USA}

\author[0000-0003-4526-3747]{Andreas Seifahrt}
\affiliation{Department of Astronomy \& Astrophysics, University of Chicago, Chicago, IL 60637, USA}

\author[0000-0003-0534-6388]{David Kasper}
\affiliation{Department of Astronomy \& Astrophysics, University of Chicago, Chicago, IL 60637, USA}

\author[0000-0002-4671-2957]{Rafael Luque}
\affiliation{Department of Astronomy \& Astrophysics, University of Chicago, Chicago, IL 60637, USA}

\author[0000-0001-7409-5688]{Guðmundur Stefánsson} 
\affil{Department of Astrophysical Sciences, Princeton University, 4 Ivy Lane, Princeton, NJ 08540, USA}
\affil{NASA Sagan Fellow}
\affil{Anton Pannekoek Institute for Astronomy, University of Amsterdam, Science Park 904, 1098 XH Amsterdam, The Netherlands} 

\author[0000-0002-4410-4712]{Julian St{\"u}rmer}
\affiliation{Landessternwarte, Zentrum f{\"u}r Astronomie der Universität Heidelberg, K{\"o}nigstuhl 12, D-69117 Heidelberg, Germany}

\author[0000-0002-9003-484X]{David~Charbonneau}
\affiliation{Harvard-Smithsonian Center for Astrophysics, 60 Garden St, Cambridge, MA 02138, USA}

\author[0000-0001-6588-9574]{Karen A.\ Collins}
\affiliation{Center for Astrophysics \textbar \ Harvard \& Smithsonian, 60 Garden Street, Cambridge, MA 02138, USA}

\author{John~P.~Doty}
\affiliation{Noqsi Aerospace Ltd., 15 Blanchard Avenue, Billerica, MA 01821, USA}

\author[0000-0002-2482-0180]{Zahra~Essack}
\affiliation{Department of Physics and Astronomy, The University of New Mexico, 210 Yale Blvd NE, Albuquerque, NM 87106, USA}

\author[0000-0002-4909-5763]{Akihiko Fukui}
\affiliation{Komaba Institute for Science, The University of Tokyo, 3-8-1 Komaba, Meguro, Tokyo 153-8902, Japan}
\affiliation{Instituto de Astrofisica de Canarias (IAC), 38205 La Laguna, Tenerife, Spain}

\author[0000-0001-9927-7269]{Ferran Grau Horta}
\affiliation{Observatori de Ca l'Ou, Carrer de dalt 18, Sant Martí Sesgueioles 08282, Barcelona, Spain}

\author{Christina Hedges}
\affiliation{NASA Goddard Space Flight Center, 8800 Greenbelt Rd, Greenbelt, MD 20771, USA}

\author[0000-0002-3439-1439]{Coel Hellier}
\affiliation{Astrophysics Group, Keele University, Staffordshire, ST5 5BG, UK}

\author[0000-0002-4715-9460]{Jon~M.~Jenkins}
\affiliation{NASA Ames Research Center, Moffett Field, CA 94035, USA}

\author[0000-0001-8511-2981]{Norio Narita}
\affiliation{Komaba Institute for Science, The University of Tokyo, 3-8-1 Komaba, Meguro, Tokyo 153-8902, Japan}
\affiliation{Astrobiology Center, 2-21-1 Osawa, Mitaka, Tokyo 181-8588, Japan}
\affiliation{Instituto de Astrofisica de Canarias (IAC), 38205 La Laguna, Tenerife, Spain}

\author[0000-0002-8964-8377]{Samuel~N.~Quinn}
\affiliation{Center for Astrophysics \textbar \ Harvard \& Smithsonian, 60 Garden Street, Cambridge, MA 02138, USA}

\author[0000-0002-1836-3120]{Avi Shporer}
\affiliation{Department of Physics and Kavli Institute for Astrophysics and Space Research, Massachusetts Institute of Technology, Cambridge, MA 02139, USA}

\author[0000-0001-8227-1020]{Richard P. Schwarz}
\affiliation{Center for Astrophysics \textbar \ Harvard \& Smithsonian, 60 Garden Street, Cambridge, MA 02138, USA}

\author[0000-0002-6892-6948]{Sara~Seager}
\affiliation{Department of Physics and Kavli Institute for Astrophysics and Space Research, Massachusetts Institute of Technology, Cambridge, MA 02139, USA}
\affiliation{Department of Earth, Atmospheric and Planetary Sciences, Massachusetts Institute of Technology, Cambridge, MA 02139, USA}
\affiliation{Department of Aeronautics and Astronautics, MIT, 77 Massachusetts Avenue, Cambridge, MA 02139, USA}

\author[0000-0002-3481-9052]{Keivan~G.~Stassun} 
\affiliation{Department of Physics and Astronomy, Vanderbilt University, Nashville, TN 37235, USA}
\affiliation{Department of Physics, Fisk University, Nashville, TN 37208, USA}

\author[0009-0008-5145-0446]{Stephanie Striegel}
\affiliation{SETI Institute, Mountain View, CA 94043 USA/NASA Ames Research Center, Moffett Field, CA 94035 USA}

\author[0000-0001-8621-6731]{Cristilyn N.\ Watkins}
\affiliation{Center for Astrophysics \textbar \ Harvard \& Smithsonian, 60 Garden Street, Cambridge, MA 02138, USA}

\author[0000-0002-4265-047X]{Joshua~N.~Winn}
\affiliation{Department of Astrophysical Sciences, Princeton University, 4 Ivy Lane, Princeton, NJ 08544, USA}

\author{Roberto Zambelli}
\affiliation{Società Astronomica Lunae, Castelnuovo Magra, Italy}

\begin{abstract}
M dwarf stars provide us with an ideal opportunity to study nearby small planets.  The HUMDRUM (HUnting for M Dwarf Rocky planets Using MAROON-X) survey uses the MAROON-X spectrograph, which is ideally suited to studying these stars, to measure precise masses of a volume-limited ($<\,30$\,pc) sample of transiting M dwarf planets.  TOI-1450 is a nearby (22.5\,pc) binary system containing a M3 dwarf with a roughly 3000\,K companion.  Its primary star, TOI-1450A, was identified by \tess to have a 2.04d transit signal, and is included in the HUMDRUM sample.  In this paper, we present MAROON-X radial velocities which confirm the planetary nature of this signal and measure its mass at a nearly 10\% precision.  The 2.04d planet, TOI-1450Ab, has $R_b\,=\,1.13\,\pm\,0.04\,R_\oplus$ and $M_b\,=\,1.26\,\pm\,0.13\,M_\oplus$.  It is the second-lowest-mass transiting planet with a high-precision RV mass measurement.  With this mass and radius, the planet's mean density is compatible with an Earth-like composition.  Given its short orbital period and slightly sub-Earth density, it may be amenable to \textit{JWST} follow-up to test whether the planet has retained an atmosphere despite extreme heating from the nearby star.  We also discover a non-transiting planet in the system with a period of 5.07\,days and a $M\mathrm{sin}i_c\,=\,1.53\,\pm\,0.18\,M_\oplus$.  We also find a 2.01d signal present in the systems's \tess photometry that likely corresponds to the rotation period of TOI-1450A's binary companion, TOI-1450B.  TOI-1450A, meanwhile, appears to have a rotation period of approximately 40\,days, which is in-line with our expectations for a mid-M dwarf.
\end{abstract}


\section{Introduction}
\label{sec:intro}

M dwarfs are ideal targets for the discovery of exoplanets, as many detection methods have signal amplitudes inversely proportional to the size of the host star.  Atmospheric characterization surveys, using instruments like \textit{JWST}, are similarly well-suited to small stars.  This fact, combined with the extreme abundance of M dwarfs in the universe, means that we expect to find a large number of small, rocky planets around M dwarfs, and that the compositions, atmospheres, and formation histories of these planets should be relatively easy to study compared to similar planets around Sun-like stars.  


Predictions of the single-transit detection capabilities of \jwst found that the majority of rocky planets with detectable atmospheric molecules orbit stars within 10 pc and with effective temperatures $T_{\mathrm{eff}}\,<\,3400$\,K \citep{Wunderlich19}.  This highlights the importance of precisely measuring the properties of M dwarf planets in our solar neighborhood.  While longer campaigns would allow us to target warmer and more distant stars, these targets would be far more resource-intensive. 

While it is relatively easy to study M dwarf planets, they may differ substantially from planets around hotter stars.  Firstly, M dwarfs tend to have much longer active periods than other types of stars, and possess more frequent and more energetic flare activity during these periods \citep[see, e.g.,][for more details]{Hawley00}.  These events could contribute to the vaporization of any surface oceans, resulting in substantial water loss as the water dissociates in the atmosphere and the hydrogen escapes to space \citep{Luger15}. Vaporized oceans could contribute to a strong abiotic O$_2$ signature, which could be detected with \jwst \citep{Fauchez20} and tell us whether or not a planet's surface is dessicated.  However, these signatures may be difficult to detect, as the high stellar activity and flare rate of M dwarfs could contribute to rapid planet atmosphere escape compared to planets at equivalent instellations around more luminous stars \citep{Tilley19}.  As escape results in thinner atmospheres with higher mean molecular weights, M dwarf planetary atmospheres may be difficult to characterize with instruments like \textit{JWST}.

Additionally, due to the dramatic lowering of an M dwarf's luminosity as it moves onto the main sequence \citep{Baraffe15}, planets' instellations change drastically over their lifetimes.  An in-situ planet around a M dwarf likely received a much larger amount of radiation in the past than it does today.  As the relative location of the snow line during planet formation dictates what sorts of planetesimals are incorporated into a forming planet, this may mean that planets that are currently in the HZ (habitable zone) formed when the snowline was far away \citep{Mulders15}.  This would result in M dwarf planets having volatile-poor compositions with higher densities than planets at similar instellations around other types of stars.  This has important implications with respect to the expected yields of transiting atmosphere surveys, as well as the generalizability of their results. 

This situation is further complicated by the sample of M dwarf planets studied in \cite{Luque2022}, which found that low-radius M dwarf planets can largely be split into two populations, and interpreted those populations as rocky planets (like Earth) and planets that are 50\% water by mass.  It is unlikely for these water-rich planets to form so close to the host star, so these strange targets may be planets that formed beyond the ice line \citep[where the rock:volatile ratio is about 1:1, see][]{Thiabaud2014, Marboef2014} and migrated inwards towards their host star \citep[which has been shown to be possible in the population synthesis models by][]{Burn2021}.  However, \cite{Rogers2023} have shown that it is possible for a similar population partitioning to be caused by a divide between rocky cores and planets with massive hydrogen atmospheres, which is the more familiar explanation for the well-known exoplanet radius gap \citep{Fulton2017}.  Understanding which of these scenarios is more likely could have a dramatic impact on our understanding of M dwarf planet formation and habitability.

To evaluate these hypotheses, we need to measure the compositions of a large, relatively unbiased sample of M dwarf planets.  As planets of different compositions can have very similar masses and radii \citep[see, e.g., the mass-radius models included in][]{Zeng19}, it is vital to collect high-precision masses and radii for nearby M dwarf planets with wide-field and/or ground-based instruments.  This will allow us to properly prioritize targets for follow-up with expensive single-target instruments like \textit{JWST}.  Precise mass measurements will also be helpful for atmospheric retrievals with \textit{JWST}, as \cite{Batalha19} found that a 20\% planetary mass precision is necessary to perform precise planetary atmospheric retrievals.  With these mass measurements, we will be able to understand whether or not M dwarf planets tend to differ systematically than planets around FGK stars.

While many nearby M dwarfs have transiting planets with 5\% precision radii measured with wide-field photometric surveys like \kepler \citep{Borucki10} and \tess \citep{Ricker15}, there is no accompanying radial velocity (RV) survey that seeks to provide every single one of these planets with precise masses.  Thus, we are conducting HUMDRUM (HUnting for M Dwarf Rocky planets Using MAROON-X), a volume-limited survey of nearby ($d\,<\,30\,pc$) M dwarf ($T_{\mathrm{eff}}$\,<\,4000\,K) planets with transits identified via \tess with MAROON-X.  By selecting targets that \tess has identified, the biases of \tess can be estimated \citep[using simulations such as][or with more advanced methods]{Sullivan2015, Brady2022} to inform the biases of this survey.  The completeness of this survey will be discussed in greater detail in following papers.  

In this paper, we present RV measurements of one of the targets in the 30 pc sample, TOI-1450A, and confirm the planetary nature of its 2.04d transit signal.  We are also able to measure the $\approx\,40$d rotation period of the primary star, as well as identify a 2d signal in the \tess photometry that may be due to the rotation of TOI-1450B.  The planet's short orbital period places it well within the conservative limits for tidal locking from \cite{Barnes2017}, meaning that the planet is likely tidally locked over Gyr timescales.  As the primary star is relatively cool ($T_\mathrm{eff}\,\approx\,3400$\,K) and nearby, it is relatively easy to observe with \textit{JWST}.  Its planet is likely rocky and contributes to the growing sample of small nearby rocky planets that are prime targets for follow-up with \textit{JWST}. 

In Section~\ref{sec:system}, we describe the TOI-1450 system.  In Section~\ref{sec:methods}, we describe the observations and data used in this paper.  We characterize the stars in Section~\ref{sec:fitting} and the planets in Section~\ref{sec:fitting_planet}.  We discuss the fit system parameters in Section~\ref{sec:discussion} and provide a summary of our conclusions in Section~\ref{sec:conclusions}.

\section{The TOI-1450 System}
\label{sec:system}

TOI-1450 is a multiple star system, consisting of an A and B component separated by a projected 76\,AU on-sky \citep{Mugrauer20}.  At a distance of 22.44\,pc \citep{GaiaEDR3}, the target falls in the 30\,pc HUMDRUM sample. As described by the \tess catalog \citep{Stassun19}, TOI-1450A is a $R_\star\,=\,0.474\,\Rsun$ M dwarf with an effective temperature of $T_{\mathrm{eff}}\,=\,3407\,\pm\,157$\,K.  TOI-1450B is described by \cite{Mugrauer20} as an M dwarf with a temperature of around 3000\,K.  Relations from \cite{Giovinazzi22} indicate that, based on its Gaia RP magnitude, TOI-1450B has a mass of around $M_*\,=\,0.14\,\pm\,0.01\,\Msun$.  TOI-1450A and TOI-1450B are separated by 3.4 arcseconds and are thus easily resolved by MAROON-X \citep{Seifahrt18}, but not by \textit{TESS}.  The two stars are close enough on-sky that TOI-1450B is not identified separately in the 2MASS catalogue \citep{2MASS, Skrutskie2006}, meaning that the observed $JHK$ magnitudes of the system are likely influenced by blending.  Thus, it is important to consider the influence of TOI-1450B when evaluating any photometric signals.  The stellar parameters of TOI-1450A are provided in Table~\ref{tab:host}.

\begin{table}[]
    \centering
    \begin{tabular}{|l|c|c|}
    \hline
    \textbf{Property} & \textbf{Value} & \textbf{Reference} \\
    \hline
    RA &  19:07:24.83 & \cite{GaiaEDR3} \\
    Declination & +59:05:09.36 & \cite{GaiaEDR3} \\
    Spectral Type & M3.0V & \cite{Lepine13} \\
    $J$ Mag & 8.457 & \cite{2MASS} \\
    $H$ Mag & 7.820 & \cite{2MASS}\\
    $K$ Mag & 7.565 & \cite{2MASS}\\
    M$_{\star,~K_s}$ (\Msun) & 0.471 $\pm$ 0.020 & \cite{Stassun19}\\
    R$_{\star,~K_s}$ (\Rsun) & 0.474 $\pm$ 0.014 & \cite{Stassun19}\\
    L$_\star$ (\Lsun) & 0.0027 $\pm$ 0.007 & \cite{Stassun19}\\
    M$_{\star,~\mathrm{SED}}$ (\Msun) & 0.480 $\pm$ 0.024 & This work \\
    R$_{\star,~\mathrm{SED}}$ (\Rsun) & 0.483 $\pm$ 0.025 & This work \\
    Distance (pc) & 22.443 $\pm$ 0.045 & \cite{GaiaEDR3} \\
    P$_{\mathrm{rot}}$ (d) & 30-40 & This work \\
    T$_\mathrm{eff}$ (K) & 3437 $\pm$ 86 & This work \\
    {[Fe/H]} & -0.12 $\pm$ 0.17 & This work \\
    log~$g$ & 4.76 $\pm$ 0.04 & This work \\
    \hline
    \end{tabular}
    \caption{Properties of TOI-1450A, the host star.}
    \label{tab:host}
\end{table}

\section{Observations}
\label{sec:methods}

\subsection{MAROON-X Radial Velocities}
\label{ssec:maroonx}

MAROON-X \citep{Seifahrt18, Seifahrt22} is a high-precision echelle spectrograph mounted on the 8.1-m telescope Gemini-North.  It has a wavelength coverage in the visible and near infrared encompassing 500\,---\,920\,nm, selected due to the expected density of M dwarf spectral lines in this regime.  The instrument has two CCDs, one ``blue'' (500\,---\,670\,nm) and the other ``red'' (650\,---\,920\,nm), which are both exposed simultaneously whenever a target is observed.  These two channels are treated as separate instruments for the purposes of data analysis, as they are in different wavelengths and thus capture different stellar signals.  The data is reduced with a \texttt{Python3} pipeline developed using tools that were originally developed for the CRIRES instrument \citep{Bean10} and the RVs are calculated using a version of \texttt{serval} \citep{Zechmeister20} updated to work  with MAROON-X data.  RVs are found by stacking the individual spectra to form a co-added template, which is then compared to the individual observations using least-squares fitting to find the individual exposure radial velocities. Telluric lines from Earth's atmosphere are fully masked to prevent them from skewing our analysis.  MAROON-X observations are calibrated with etalon spectra, which are observed simultaneously with the science targets. The etalon spectra correct for velocity drifts of the spectrograph but are themselves prone to long-term drifts due to aging of the Zerodur spacer in the etalon optics. The long-term drifts as well as additional offsets between runs were calibrated out using the spectra of a ThAr lamp.

\begin{figure}
    \centering
    \includegraphics[width=0.9\linewidth]{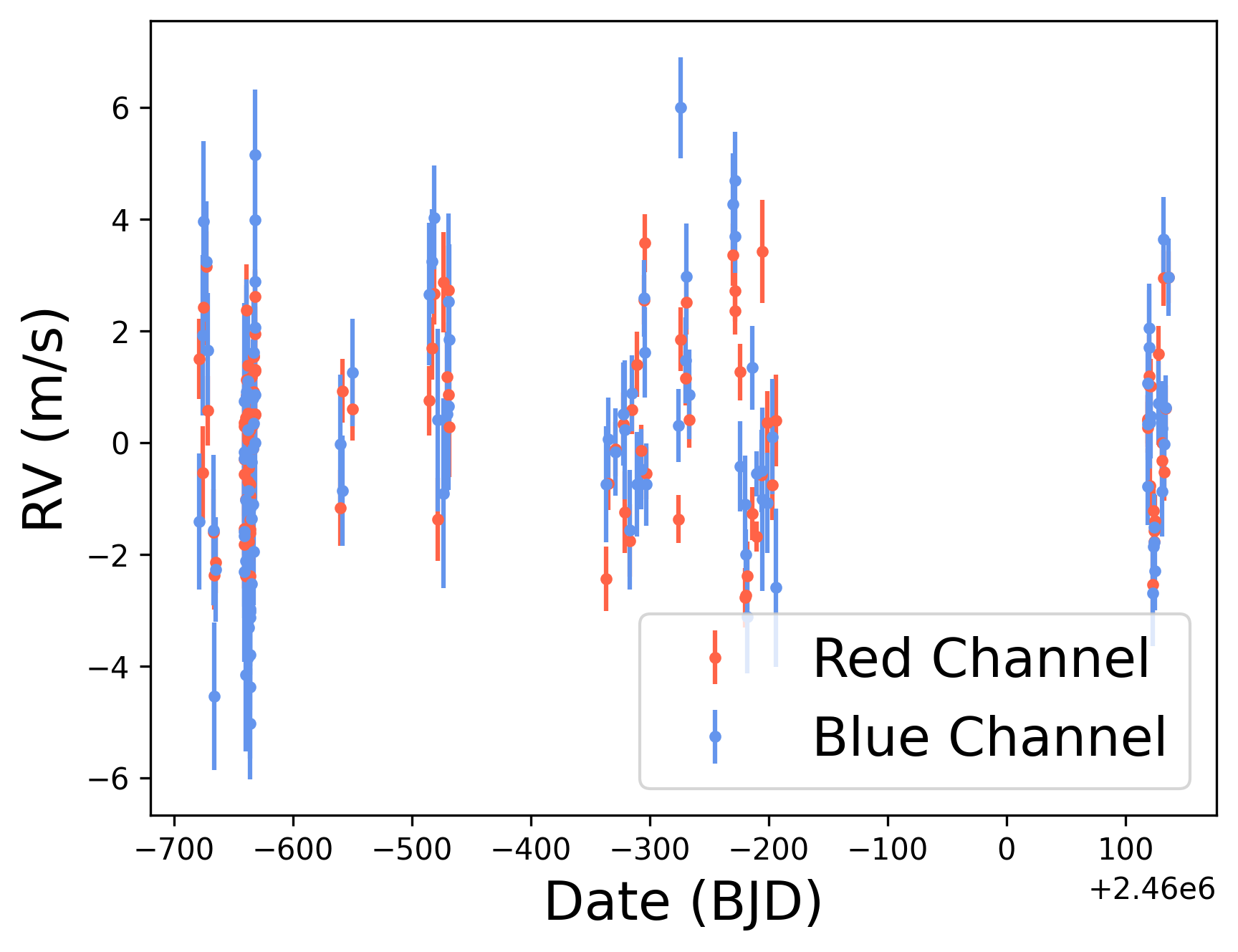}
    \caption{The MAROON-X RV observations of TOI-1450 with calibrated offsets between runs applied when available (and mean-subtractions otherwise). The red-channel data are plotted in red, while the blue-channel data are plotted in blue.}
    \label{fig:toi1450-rv}
\end{figure}

We observed TOI-1450A as a part of the HUMDRUM sample, observing the system 118 times between the months of April 2021 and July 2023.  As the planet has an orbital period very close to 2 days, we attempted to observe it multiple times a night to improve our phase coverage.  The data are shown in Figure~\ref{fig:toi1450-rv}.  Most exposures, observed simultaneously in the red and blue channels, were ten minutes long, allowing for peak per-pixel SNRs of approximately 80 in the blue channel and 200 in the red channel.  The higher signal in the red channel is expected due to the cool nature of the star. These peak SNRs translate to a per-measurement observed RV error of roughly 1.0\,m/s in the blue channel and 0.6\,m/s in the red.

Over the course of our observations, MAROON-X was operated in campaign mode on Gemini-North, sharing a port with several other instruments.  Thus, the target was observed over the course of eight runs, discrete periods in which the instrument was connected to the telescope.  There are frequently small (on the order of a few m/s or less) RV offsets between these runs.  These offsets are the result of both a constant, linear drift in our etalon calibrations over time (due to the aging of the ThAr lamp) in addition to some smaller, random instrumental shifts that are on the order of 1-2\,m/s and occur between each run, likely as the result of connecting and disconnecting the fiber optic feed of the instrument from the telescope.  Major instrumental failures or interruptions, such as the major cooler system shutdown on the instrument between April and May 2021, also imparted additional offsets.  These offsets were calibrated by the examination of RV standard stars, such as HD 88230, GJ 908, HD 3651, and HD 32147, during each MAROON-X run.  Offsets between each MAROON-X run were then calculated using the measured RVs of these stars.  These offsets (listed in Table~\ref{tab:priors_joint}) typically had errors on the order of 0.5\,m/s, which we included as priors when fitting the RV data.  We also included the effects of an observed linear RV drift observed in all MAROON-X targets in August 2021, calibrated using several standard stars.  As there were only three observations in August, any error on this drift calibration is unlikely to meaningfully affect our results.

We also observed TOI-1450B with MAROON-X twice in July 2022, in an effort to characterize the star.  As TOI-1450B is much dimmer than TOI-1450A, we performed 30 minute exposures when observing this target.  The resulting spectra had per-pixel SNRs of around 20 in the blue channel and 70 in the red channel. This data is insufficient to do a full characterization of the star's orbit but give us some insight into its spectral characteristics.

Our RVs are included in Table~\ref{tab:RVs}.

\begin{table}
\begin{filecontents*}{TOI1450_RVs_truncated.csv}
t,channel,run,rv,erv,crx,ecrx,ha,eha
2321.07733,red,Apr21,-5.89,0.72,17.44,10.77,0.61022,0.00433
2321.07733,blue,Apr21,-8.48,1.22,18.59,18.21,0.58238,0.00373
2324.09946,red,Apr21,-7.93,0.83,12.59,9.2,0.61664,0.00547
2324.09946,blue,Apr21,-5.15,1.44,15.96,24.91,0.60315,0.00463
2324.99367,red,Apr21,-4.97,0.68,-4.11,12.68,0.68869,0.00444
2324.99367,blue,Apr21,-3.1,1.44,9.66,28.29,0.65631,0.00386
\end{filecontents*}
\csvreader[centered tabular= |c|c|c|c|,
    table head=\hline Time (BTJD) & Channel & RV (m/s) & RV Error \\\hline,
    late after line=\\, 
    late after last line = \\\hline,
    ]
{TOI1450_RVs_truncated.csv}{t=\t, channel=\channel, rv=\rv, erv=\erv }
{ \t & \channel & \rv &\erv }
\caption{The first several MAROON-X RVs collected on TOI-1450A. The times are in BTJD (BJD\,-\,2,457,000).  The rest of the RVs, as well as the values of the activity indicators, are available online. \label{tab:RVs}}
\end{table}

\subsection{TESS Photometry}
\label{ssec:tess}

The Transiting Exoplanet Survey Satellite \citep[\tess,][]{Ricker15} is a wide-field all-sky photometric survey which observes large chunks of the sky in 27-day sectors.  

The transiting signal around TOI 1450 was originally identified after the SPOC conducted a transit search of Sector 14 with an adaptive, noise-compensating matched filter \citep{Jenkins2002, Jenkins2010, Jenkins2020} identifying a threshold-crossing event (TCE) for which an initial limb-darkened transit model was fitted \citep{Li2019} and a suite of diagnostic tests were conducted to help determine the planetary nature of the signal \citep{Twicken2018}. The transit signature was also detected in a search of Full Frame Image (FFI) data by the Quick Look Pipeline (QLP) at MIT \citep{Huang2020a, Huang2020b}. The TESS Science Office (TSO) reviewed the vetting information and issued an alert on 14 November 2019 \citep{Guerrero2021}. The signal was repeatedly recovered as additional observations were made and the transit signature passed all the diagnostic tests presented in the Data Validation reports. According to the difference image centroiding tests, the host star is located within 5.95$\pm$\,3.48 arcseconds of the source of the transit signal.

TOI-1450 has been observed in 27 sectors (14-17, 19-26, 40-41, 47, 49-60), and the data encompass dates from July 2019 through January 2023.  This is substantially more data than is available for most other \tess targets, which typically have only a few sectors' worth of data available.  The data are available at a two-minute cadence, much faster than the thirty-minute cadence full frame images (FFIs).  Starting in sector 40, as a part of the \tess extended mission, the data are available with a 20-second cadence, allowing for further refinements in our understanding of the transit shape.  The data from sectors 49-54 overlap with MAROON-X observations.  All image data were reduced and analyzed by the TESS Science Processing Operations Center at NASA Ames Research Center \citep{Jenkins16}.

Given the large pixels of TESS (21 arcseconds), we checked the target field to estimate how much the transit signal was diluted by light from other stars and to ensure that we targeted the correct star for follow-up.  The presence of nearby bright stars can result in smaller planetary transits and thus an underestimate of the planet's radius.

\begin{figure}
    \centering
    \includegraphics[width=0.9\linewidth]{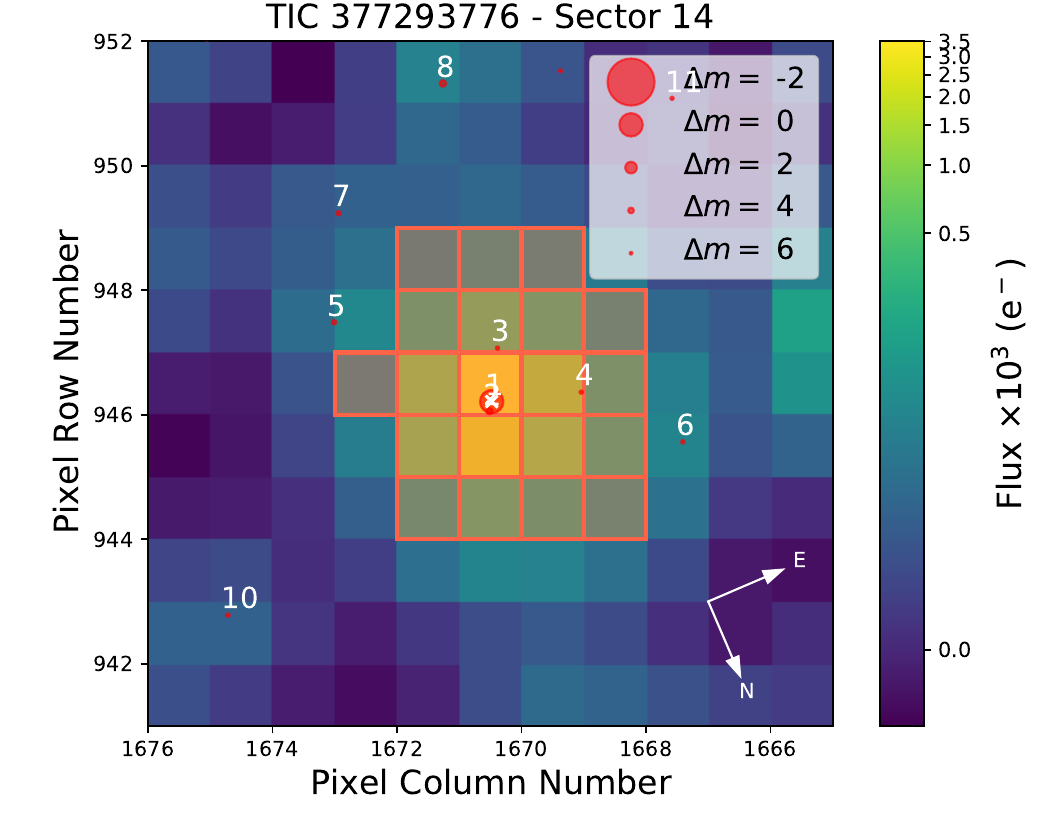}
    \caption{The crowding around TOI-1450.  The sky, as seen by \tess in Sector 14, is shown, with the specific aperture mask used marked in red.  The star studied in this paper is marked with a white X, while nearby stars (identified via GAIA) are shown as red circles, with their size representative of their $G$ magnitude relative to TOI-1450A. Only stars with a $\Delta m_G\,\leq\,6$ are shown, as the brightest stars contribute maximally to the light curve flux contamination.}
    \label{fig:crowding}
\end{figure}

Investigating the crowding is fairly simple using the code \texttt{tpfplotter} \citep{Aller20}, which plots the location of the \tess target, the \tess aperture mask, and the locations of any nearby stars found via the GAIA EDR3 \citep{GaiaEDR3}.  Figure~\ref{fig:crowding} shows the stars that have a chance of causing flux contamination in the Sector 14 data.  It is obvious that most of the stars that fall within the TOI-1450 aperture are substantially dimmer than TOI-1450A, with the brightest other star (TOI-1450B) having a $\Delta m_G\,=\,3.78$.

We can determine the precise amount of dilution from the B star in the \textit{TESS} light curve by estimating its $T$ magnitude $m_T$.  Using the relationship between the Gaia colors and $m_T$ from \cite{Stassun19}, we find that TOI-1450B has $m_T\,=\,13.70$.  This means that $\Delta m_T\,=\,3.67$, corresponding to a dilution of around 3.3\%.  Even when also considering the other stars in the aperture, the dilution is less than 5\%.  If the transit is occurring around the B star, the dilution is $>\,90\%$, meaning that the transit appears to be much shallower than it actually is.  However, our MAROON-X RVs confirm that the 2d transiting planet is orbiting around the primary star (see Section~\ref{ssec:rv}). Thus, the corrections we need to make to the light curve due to dilution are minimal, and are already performed by the Presearch Data Conditioning SAP (PDCSAP) algorithm \citep[for more details on how crowding is dealt with by the PDCSAP algorithm, see][]{Stumpe2012, Smith2012, Stumpe2014}.

\subsection{ASAS Photometry}

\begin{figure}
      \centering
		\includegraphics[width=\linewidth]{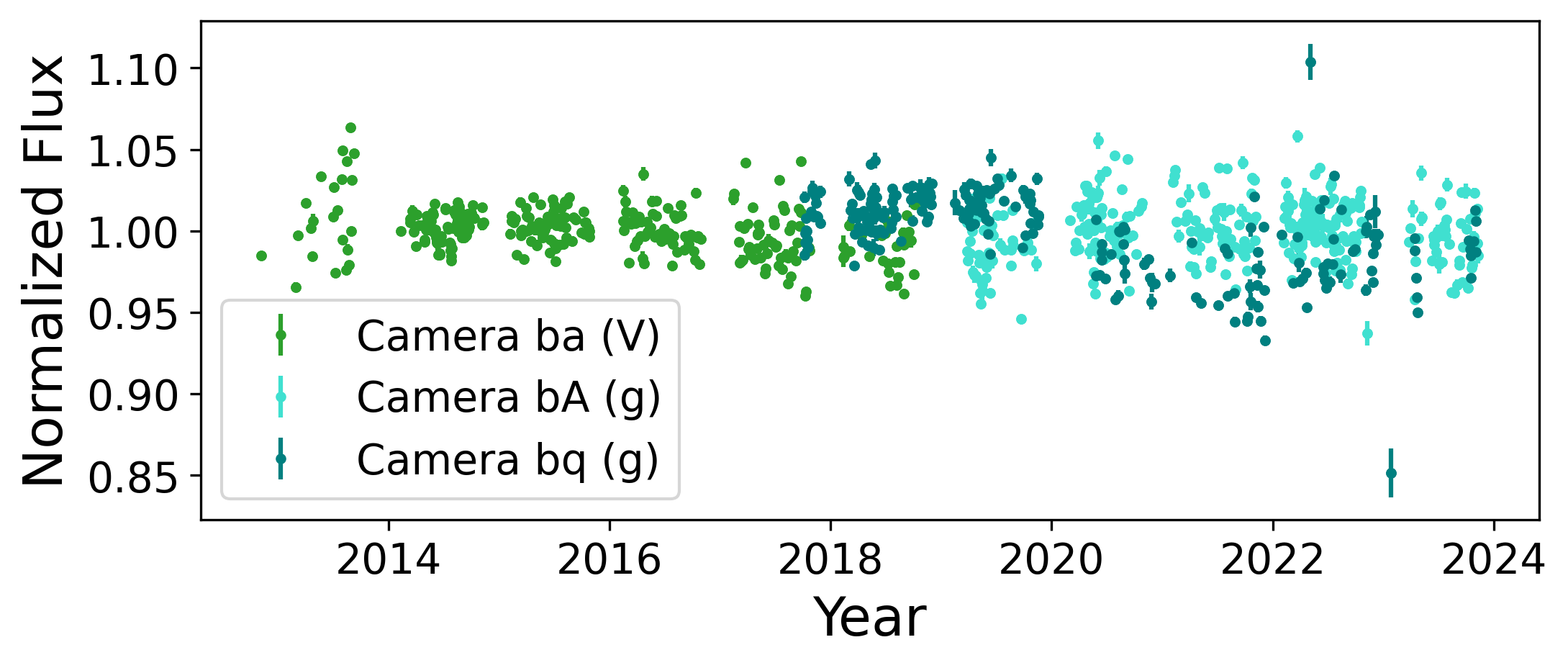}
	\vfill
		 \includegraphics[width=\linewidth]{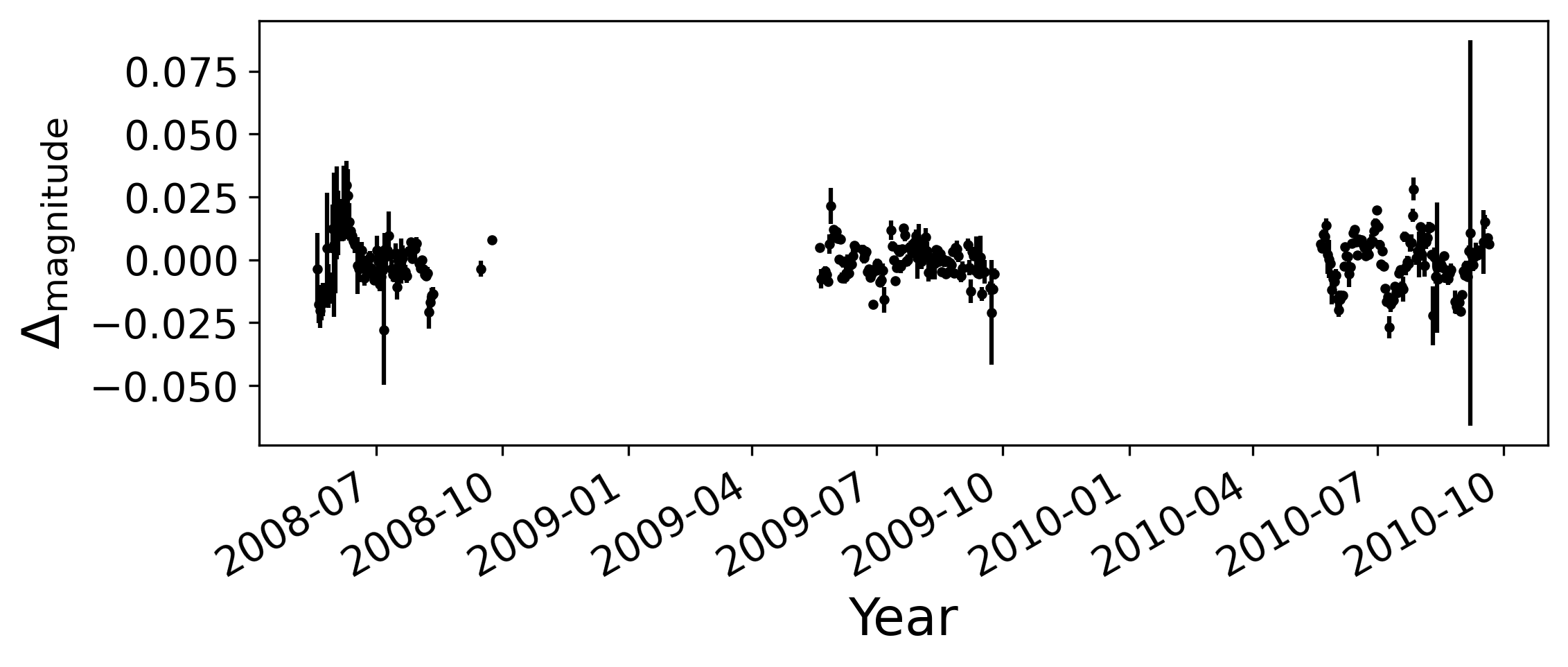}
	\caption{Top: ASAS-SN $V$ and $g$ band photometry of TOI-1450, normalized to unit mean flux.  Bottom: 1d-binned WASP photometry of TOI-1450.}
	\label{fig:other_lc}
\end{figure}

The TOI-1450 system has also been observed with ASAS-SN \citep{Shappee14, Kochanek17}, an all-sky photometric survey.  The photometry, as of 11/14/2023, is shown in Figure~\ref{fig:other_lc}.  Given ASAS-SN's precise calibrations and long time baseline, it is a better photometer for examining long-term signals in the TOI-1450 light curve than \textit{TESS}.  However, as ASAS-SN images typically have cadences on the order of days (while the transits have one-hour durations), they are not useful for detecting the planetary transits.  ASAS-SN images have larger pixels (8\,arcseconds) than the separation between TOI-1450A and TOI-1450B, meaning that the observed light curves are the result of the blended light of the two targets.  Given how much dimmer TOI-1450B is than TOI-1450A, however, we can expect TOI-1450A to be the dominant source of flux in the ASAS-SN curve.

TOI-1450 has flux measurements in $V$ and $g$ (filters centered on 551 and 480\,nm, respectively) available in the ASAS-SN public database\footnote{https://asas-sn.osu.edu/}.  We downloaded flux measurements from the online database, processing them using the publicly available image subtraction photometry pipeline, which performs photometry on coadded, image subtracted data.  As these light curves use coadded data, they contain fewer individual data points but are less noisy than the pure aperture photometry light curves.  There were 309 total points in the $V$ band and 572 total measurements in the $g$ band. 

The $V$ band data are older, encompassing times from November 2012 to October 2018, with an median observation rate of once per every three days.  The $g$ data are the result of an overall instrument overhaul and are both more recent (October 2017 to the present day) and have a more rapid cadence (with a median cadence of one observation every two days).

\subsection{LCOGT Photometry}

The \textit{TESS} pixel scale is $\sim 21\arcsec$ pixel$^{-1}$ and photometric apertures typically extend out to roughly 1 arcminute, generally causing multiple stars to blend in the \textit{TESS} photometric aperture. To attempt to determine the true source of the \textit{TESS} detection, we acquired ground-based time-series follow-up photometry of the field around TOI-1450A as part of the \textit{TESS} Follow-up Observing Program \citep[TFOP;][]{collins:2019}\footnote{https://tess.mit.edu/followup}.

We observed two full transit windows of TOI-1450.01 on UTC 2022 July 6 and UTC 2022 August 17 in Pan-STARRS $z$-short band from the the Las Cumbres Observatory Global Telescope \citep[LCOGT;][]{Brown:2013} 1\,m network nodes at Teide Observatory on the island of Tenerife (TEID) and McDonald Observatory near Fort Davis, Texas, United States (McD), respectively. We also observed one full transit window on UTC 2022 September 03 simultaneously in Sloan $g'$, $r'$, $i'$, and Pan-STARRS $z$-short from the LCOGT 2\,m Faulkes Telescope North at Haleakala Observatory on Maui, Hawai'i. The Faulkes Telescope North is equipped with the MuSCAT3 multi-band imager \citep{Narita:2020}. The images were calibrated by the standard LCOGT {\tt BANZAI} pipeline \citep{McCully:2018} and differential photometric data were extracted using {\tt AstroImageJ} \citep{Collins:2017}. We used circular photometric apertures centered on TOI-1450A with radius $5\farcs8$ for the 1\,m observations and $4\farcs9$ for the MuSCAT3 observation to extract the on-target differential photometry. The nearest known neighbor in the Gaia DR3 catalog (Gaia DR3 2155540255030594688; TOI-1450B) is $\sim 3\farcs3$ north of TOI-1450A and is blended in the apertures. The light curves are included in the joint modeling described in Section \ref{ssec:joint}. Using smaller $\sim3\arcsec$ photometric apertures, we tentatively detected the $\sim0.5$\,ppt transit in TOI-1450A, but with higher scatter and systematics in the light curves. We also ruled out nearby eclipsing binaries (NEBs) within 2.5$\arcmin$ of TOI-1450A, except for the $\sim3\farcs2$ neighbor TOI-1450B, which is heavily contaminated by the $\sim30$ times brighter target star.

\subsection{WASP Photometry}

TOI-1450 was also observed by the WASP project at the Observatorio del Roque de los Muchachos \citep{Pollacco2006} between the years of 2008 and 2010.  The photometry is shown in Figure~\ref{fig:other_lc}.  The WASP transit-search operated arrays of 8 cameras using 200mm, f/1.8 lenses with a broadband filter spanning 400--700 nm, backed by 2048x2048 CCDs giving a plate scale of $13.7\arcsec$/pixel \citep{Pollacco2006}.  The large pixel size means that the WASP data also features blending between TOI-1450A and TOI-1450B on a single pixel.  TOI-1450 was observed over spans of $\sim$\,120 days in each year. Observations on every clear night, with a typical 15-min cadence, accumulated 27\,600 photometric data points. 

\section{Analysis}
\label{sec:fitting}

\subsection{Stellar Parameter Fits}

\subsubsection{Direct Comparison to PHOENIX Spectra}

As we currently lack an interferometic measurement of the star's radius, the parameters of TOI-1450 are those from the \tess catalog \citep{Stassun19}, which takes its parameters from \cite{Muirhead18}, which makes use of the relations from \cite{Mann15} and \cite{Mann19} to estimate the stellar characteristics. 

We can provide an independent measurement of the stellar parameters by comparing the MAROON-X spectra directly to stellar atmospheric models, allowing us to find the $T_\mathrm{eff}$, log~$g$, and [Fe/H] of TOI-1450 directly.  We thus produced a code, with a similar methodology as that from \cite{Passegger18}, which estimates the stellar parameters by comparing the gathered MAROON-X spectrum to a set of interpolated, broadened \texttt{PHOENIX} \citep{Husser13} models.  We consider both rotational broadening and the derived instrument broadening order-by-order of MAROON-X, which can be inferred directly from the etalon calibration files.  

Instead of fitting a model to the entire MAROON-X spectrum, we have decided to primarily focus on regions that have shown to be sensitive to the parameters of interest, inspired by those used by CARMENES \citep{Passegger18}, which is an instrument with very similar wavelength coverage to MAROON-X.  Avoiding a full spectral comparison both makes our algorithm run substantially faster and allows us to focus primarily on the regions that are known to be the most indicative and/or modelled accurately.  We have removed wavelength regions that are not typically observed by MAROON-X and have avoided including regions where the model spectra typically are very poor fits to the observed data. Table~\ref{tab:regions} indicates the precise wavelengths of these regions, as well as the molecular lines these regions typically encompass.  We note that we have removed the K~I region from our analysis, as we found that the \texttt{PHOENIX} models were poor fits to this region in our models.

\begin{center}
\begin{table}[]
\begin{tabular}{|l|l|l|}
\hline
Line       & $\lambda_\mathrm{min}$ & $\lambda_\mathrm{max}$ \\
\hline
TiO        & 7055.0                 & 7074.0                 \\
TiO        & 7085.0                 & 7114.0                 \\
TiO, Ca I  & 7125.0                 & 7170.0                 \\
Ti I       & 8414.3                 & 8415.0                 \\
Ti I       & 8428.2                 & 8429.5                 \\
Ti I       & 8436.8                 & 8438.7                 \\
Ti I, Fe I & 8470.2                 & 8471.1                 \\
Fe I       & 8516.0                 & 8517.0                 \\
Fe I, Ti I & 8676.5                 & 8678.2                 \\
Ti I       & 8684.5                 & 8685.8                 \\
Mg I       & 8808.7                 & 8809.7                 \\
Fe I       & 8826.0                 & 8827.5                 \\
\hline
\end{tabular}
\caption{The rest-frame wavelengths of the regions considered in our spectral analysis, inspired by \citep{Passegger18} and slightly modified to include only regions where the models don't show strong systematic deviations from the data.}
\label{tab:regions}
\end{table}
\end{center}

The \texttt{PHOENIX} models from \cite{Husser13} have varying temperatures, surface gravities, metallicities, and $\alpha$ element abundances.  For the sake of our fits, we included models with $4\,<\,$log~$g\,<6\,$, $2300\,<\,T_\mathrm{eff}\,<\,5000$ K, and $-1.0\,<\,[Fe/H]\,<\,1.0$.  We allow the $\alpha$ abundances of the models to remain fixed at the solar values, as \texttt{PHOENIX} models with nonzero $\alpha$ values are not available for stars under 3500\,K.  These limits are guided both by the the types of stars in our sample (M dwarfs) and the actual \texttt{PHOENIX} models available for download.   We have found that all of the stars in the HUMDRUM sample fall well within these limits (according to the \tess parameter estimates).  

We also applied rotational and instrumental broadening to these models before performing any fits.  We first used the  rotational convolution kernel from \cite{Gray08}, and then applied a kernel that represents the instrumental broadening of MAROON-X (estimated from the instrument's etalon calibrations).  After this, we resampled the models according to the MAROON-X wavelengths using the \texttt{spectres} python module \citep{Carnall17}. We allowed the rotation to vary between $0\,<\,v$sin$i\,<\,5$\,km/s.

After performing broadening on each of these model spectra, we used the \texttt{scipy} module \texttt{LinearNDInterpolator} to interpolate between the model spectra, with grid spacing $\Delta T_\mathrm{eff}\,=\,100$\,K, $\Delta$log~$g$\,=\,0.5, $\Delta$[Fe/H]\,=\,0.5\,dex, and $v$sin$i$\,=\,0.5\,m/s.  The grid spacings for the first three parameters are guided by the available \texttt{PHOENIX} models, while the grid spacing of the stellar rotational velocities was user-specified and selected to be fine enough to account for the somewhat nonlinear relationship between stellar rotation and line broadening but coarse enough to allow for a reasonable computation time.  

We also fit an additional error term $\sigma$, added in quadrature to the observed spectral flux errors, in order to get a sense of the scale of the systematic differences between the models and the data.  As we normalized both the data and models such that the flux continuum was at a value of 1, this error value is expected to be between 0 and 1.  
We then performed MCMC fits of our hightest-SNR observed spectra to the interpolated model spectra using the \texttt{emcee} package, with free parameters of $T_\mathrm{eff}$, [Fe/H], $v$sin$i$, and $\sigma$.    Our interpolations are performed in their entirety before the MCMC run is initiated instead of having the broadening recalculated at every step for the sake of reducing computation time.  We did not independently fit log~$g$ because, much like \cite{Passegger18}, we found that there existed strong degeneracies between the fit surface gravity and the other parameters.  We thus avoided fitting the log~$g$ directly, and instead estimated log~$g$ at each MCMC step.  To do so, we used the estimated metallicity and the $K_s$ magnitude from the 2MASS survey \citep{2MASS} to calculate the stellar radius using the relation from \cite{Mann15} and the stellar mass using the relation from \cite{Mann19}.  \cite{Mann15} quotes a 2.7\% error on the estimated radius, while \cite{Mann19} quotes a roughly 3\% error, which overall result in our log~$g$ estimates having an additional error on the order of $\delta$log~$g \,=\, 0.025\,-\,0.030$. This systematic error will be added in quadrature to the fit error when listing the log~$g$ value.

In general, we found typical $\sigma$ values that were on the order of ten to one hundred times larger than the spectral flux errors.  This indicates that the deviations between the \texttt{PHOENIX} models and are data are dominated by systematic differences between the models and data and not merely data errors.  The fit parameter errors of this method are likely an underestimate, as there are significant discrepancies with the underlying models.  

\cite{Passegger2022} found that, even when analyzing similar spectra, different methods for stellar parameter determination can report differences in $T_\mathrm{eff}$ on the order of $\geq$ 100\,K and [Fe/H] on the order of $\approx 0.1\,$---\,0.3.  Thus, instead of calibrating our code by looking at the results of other stellar parametric determinations, we estimate the errors on our parameters by comparing them directly to data.  To estimate the accuracy of our algorithm, we compared our code's results to the known temperatures of nearby M dwarfs who have had these parameters estimated directly using interferometry.  We use these to check our algorithm's accuracy, as their parameters were determined in a manner independent of the \texttt{PHOENIX} models.  Table~\ref{tab:calibrators} lists the stars we use as references, along with their radii and temperatures.  Observations of M dwarfs in binary systems with well-characterized primary stars may be a useful way to more accurately calibrate this relationship.

\begin{table*}[]
\begin{tabular}{|l|c|c|c|c|c|}
\hline
\textbf{Name} & \textbf{Radius (R$_{\odot}$)} & \textbf{T$_\mathrm{eff}$ (K)} & \multicolumn{1}{l}{\textbf{Ref.}} & \textbf{Fit T$_\mathrm{eff}$ (K)} &  \textbf{Fit [Fe/H]} \\ \hline
GJ 15A        & 0.387 $\pm$ 0.002              & 3563 $\pm$ 11          & a & 3566$^{+13}_{-10}$ & -0.36$^{+0.04}_{-0.03}$                               \\
GJ 273        & 0.32  $\pm$ 0.005              & 3253 $\pm$ 39          & b & 3342$^{+17}_{-20}$ & -0.30$^{+0.04}_{-0.03}$                               \\
GJ 380        & 0.642 $\pm$ 0.005              & 4081 $\pm$ 15          & a & 4021$^{+17}_{-19}$ & -0.23$^{+0.03}_{-0.02}$                               \\
GJ 406        & 0.159 $\pm$ 0.006              & 2657 $\pm$ 20          & b & 3209$^{+13}_{-19}$ & 0.14$\pm 0.04$                               \\
GJ 411        & 0.387 $\pm$ 0.004              & 3547 $\pm$ 40          & a & 3519$^{+20}_{-74}$ & -0.35$^{+0.04}_{-0.15}$                               \\
GJ 447        & 0.196 $\pm$ 0.01               & 3264 $\pm$ 24          & b & 3289$^{+10}_{-11}$ & -0.25$\pm 0.03$                               \\
GJ 486        & 0.339 $\pm$ 0.015              & 3291 $\pm$ 75          & c & 3387$^{+9}_{-19}$ & -0.13$^{+0.02}_{-0.03}$                               \\
GJ 699        & 0.185 $\pm$ 0.001              & 3221 $\pm$ 32          & a & 3250$^{+9}_{-8}$ & -0.48$^{+0.03}_{-0.02}$                               \\
GJ 725A       & 0.356 $\pm$ 0.004              & 3407 $\pm$ 15          & a & 3341$^{+9}_{-8}$ & -0.49$^{+0.02}_{-0.01}$                               \\
GJ 725B       & 0.323 $\pm$ 0.006              & 3104 $\pm$ 28          & a & 3304$\pm 7$ & -0.46$\pm 0.03$                               \\
GJ 729        & 0.205 $\pm$ 0.006              & 3162 $\pm$ 30          & b & 3316$^{+9}_{-30}$ & -0.39$^{+0.03}_{-0.04}$                               \\
GJ 820A       & 0.665 $\pm$ 0.005              & 4548 $\pm$ 64          & a & 4356$^{+8}_{-11}$ & -0.42$\pm 0.02$                               \\
GJ 820B       & 0.61  $\pm$ 0.018              & 3954 $\pm$ 28          & a & 4028$^{+6}_{-4}$ & -0.32$\pm 0.01$     \\
\hline
\end{tabular}
\caption{Stars observed by MAROON-X with direct interferometric radius and temperature measurements, as well as the temperature and metallicity estimates from our code.  We note that errors associated with our fit values are purely taking the fit errors into account, and don't consider systematics.  They are thus likely systematics.  References: a) \cite{Boyajian12}, b) \cite{Rabus19}, c) \cite{Caballero22}.}
\label{tab:calibrators}
\end{table*}

\begin{figure}
    \centering
    \includegraphics[width=0.9\linewidth]{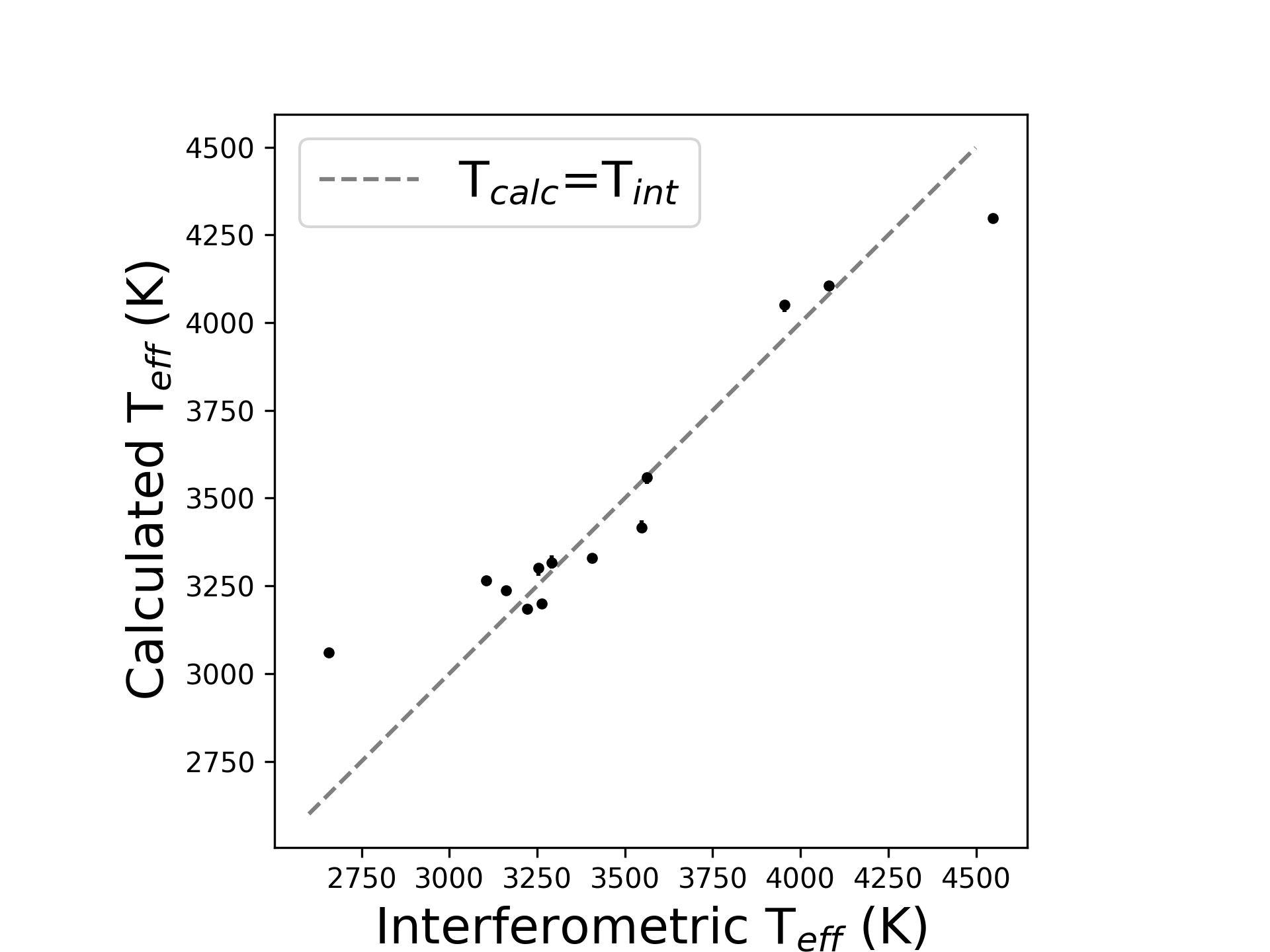}
    \caption{The interferometric temperatures of the calibrator stars compared to the MAROON-X estimations.}
    \label{fig:int_calib}
\end{figure}

Figure~\ref{fig:int_calib} compares the temperatures of our calibrators derived via interferometry with those derived using our method.  For each calibrator, we utilized the highest-SNR MAROON-X spectrum that we had available.  It is obvious that our code is less accurate at temperatures above 4200\,K and at temperatures at or below around 3200\,K.  At these temperatures, the code's sensitivity to temperature drops and the estimated temperatures appear to approach a constant value.  The lines we selected for analysis do not appear to be useful for measuring the properties of stars outside of this temperature range, which makes sense because our list was inspired by \cite{Passegger18}'s studies of M dwarfs.  While the code underestimates the temperatures for stars with temperatures above 4200\,K, such stars are not included in the HUMDRUM sample and are thus not a problem.  TOI-1450A (with an estimated temperature around 3400\,K in the \tess catalog) is unlikely to fall outside of our range of sensitivity, but TOI-1450B, which is expected to have $T_\mathrm{eff} \,\approx\, 3000$\,K, is likely too cool to obtain accurate results.

Overall, omitting Gl 406 and Gl 820A (both of which fall substantially outside of the sensitive temperature range) from our calibrator sample, we find a typical RMS deviation between our fits and the interferometric temperatures of about $\sigma_T\,=\,82$\,K.  This indicates the necessity of adding an additional error term in quadrature to the model-estimated $T_\mathrm{eff}$ (which is used to estimate log~$g$) on top of the statistical fit error and the systematic errors.  As we are using similar spectral regions as \cite{Passegger18}, we quote a fit metallicity error of $\sigma_{\mathrm{Fe/H}}\,=\,0.16$.  Unfortunately, this number may not be accurate, but we do not have a sample of M dwarfs with independently-verified metallicities to confirm our measurements.

With our code, we find that TOI-1450A has a $T_\mathrm{eff}\, = \,3433 \,\pm\, 87$\,K, log~$g$\, = \,$4.76\, \pm \,0.04$, and [Fe/H]$\,=\, -0.12 \,\pm \,0.17$.  Its metallicity is similar to that of the Sun, and may be slightly subsolar, while its temperature is exactly what we would expect for a mid-M dwarf.  

Unfortunately, TOI-1450B likely falls in the regime in which our temperature measurements are unreliable.  An effort to measure its temperature yields $T_\mathrm{eff} \, = \, 3136\,\pm\,82$\,K and its metallicity is measured as [Fe/H]$\,=\, 0.18 \,\pm \,0.17$.  Given the fact that the line list chosen seems to be unable to accurately estimate stellar parameters for stars with temperatures below about 3200\,K, these values are likely inaccurate, though it is difficult to more precisely quantify these errors without more calibrator stars.  We thus do not claim a measurement of the TOI-1450B stellar parameters.

\subsubsection{SED Fitting}

\begin{figure}
    \centering
    \includegraphics[width=0.75\linewidth,trim=100 70 60 60,clip]{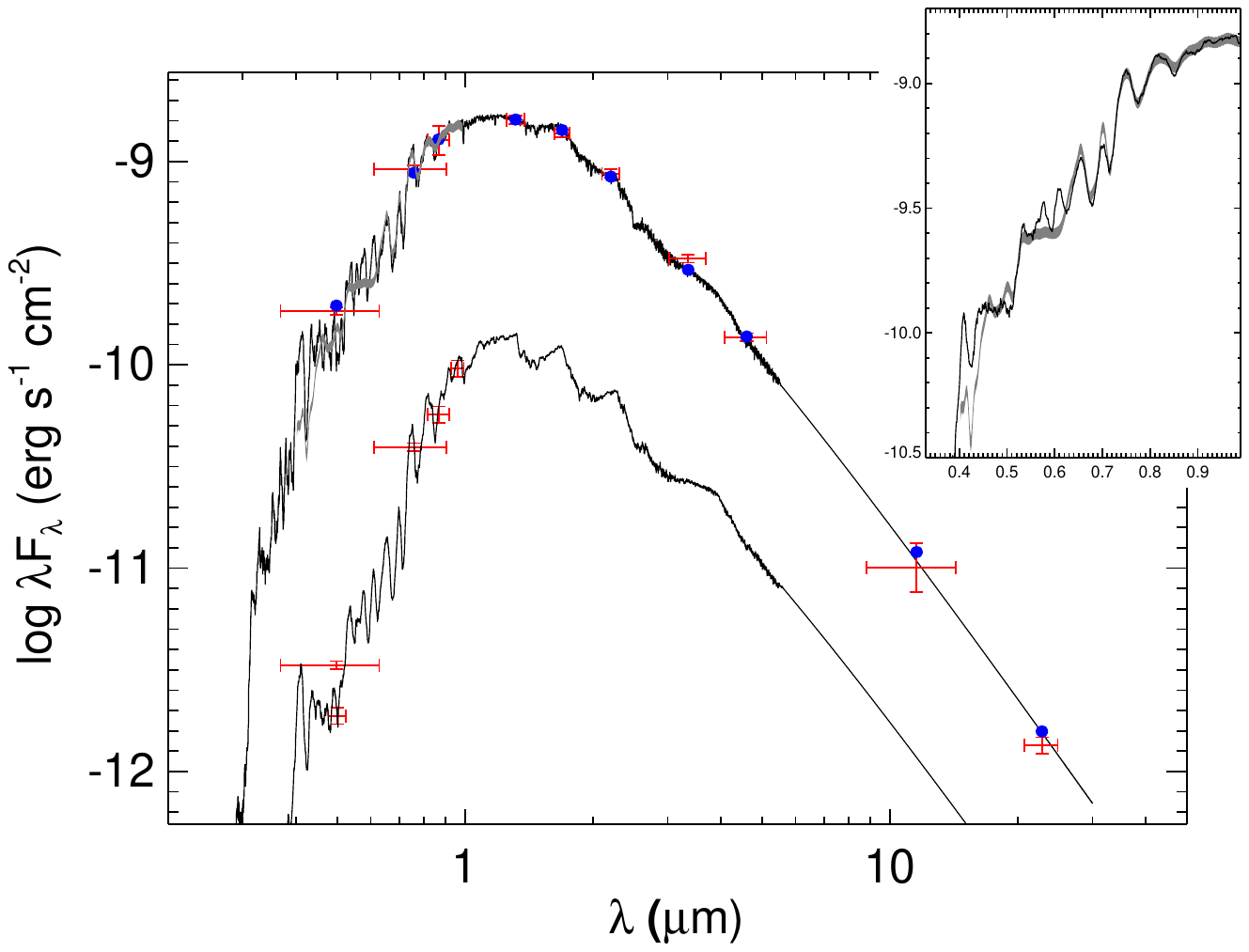}
    \caption{SED fit (upper black curve) to the observed photometry of TOI-1450A (red symbols).  The absolute flux-calibrated {\it Gaia} spectrum is shown in gray and shown in detail in the inset plot. The SED of the faint companion star, TOI-1450B, is shown as the lower black curve atop its broadband photometric measurements (red symbols).}
    \label{fig:sed}
\end{figure}

As an independent determination of the basic stellar parameters, we performed an analysis of the broadband spectral energy distribution (SED) of the star together with the {\it Gaia\/} DR3 parallax \citep[with no systematic offset applied; see, e.g.,][]{StassunTorres:2021}, in order to determine an empirical measurement of the stellar radius, following the procedures described in \citet{Stassun:2016,Stassun:2017,Stassun:2018}. We pulled the the $JHK_S$ magnitudes from {\it 2MASS}, the W1--W4 magnitudes from {\it WISE}, the $z$ magnitude from {\it PAN-STARRS}, and the $G G_{\rm BP} G_{\rm RP}$ magnitudes from {\it Gaia}. We also utilized the absolute flux-calibrated spectrophotometry from {\it Gaia}. Together, the available data spans the full stellar SED over the wavelength range 0.4--20~$\mu$m (see Figure~\ref{fig:sed}). 

We performed a fit using PHOENIX stellar atmosphere models \citep{Husser13}, with the effective temperature ($T_{\rm eff}$) and metallicity ([Fe/H]) adopted from the analysis above. The extinction $A_V$ was fixed at zero due to the close proximity of the system. The resulting fit (Figure~\ref{fig:sed}) has a reduced $\chi^2$ of 1.4. Integrating the model SED gives the bolometric flux at Earth, $F_{\rm bol} = 1.991 \pm 0.046 \times 10^{-9}$ erg~s$^{-1}$~cm$^{-2}$. Taking the $F_{\rm bol}$ and $T_{\rm eff}$ together with the {\it Gaia\/} parallax, gives the stellar radius, $R_\star = 0.500 \pm 0.026$~R$_\odot$. In addition, we can estimate the stellar mass from the empirical $M_K$ relations of \citet{Mann19}, giving $M_\star = 0.496 \pm 0.025$~M$_\odot$. 

Next, we sought to account for the small flux contribution from TOI-1450B. To do this, we pulled the available broadband photometry from catalogs in which the companion is separately resolved, namely {\it Gaia\/} and {\it PAN-STARRS}. Next, we estimated the companion star's $T_{\rm eff}$ by interpolating the relative {\it Gaia} $G$ magnitudes in the empirical tables provided in \citet{Pecaut2013}, and found that a PHOENIX model atmosphere of that $T_{\rm eff}$ (and adopting the same [Fe/H] as for TOI-1450A) indeed provides a good fit to the photometry (see Figure~\ref{fig:sed}). Finally, we computed from this SED model the companion star's contribution to $F_{\rm bol}$ overall and to the $K_S$ magnitude specifically, then recalculated the radius and mass for TOI-1450A, giving $R_\star = 0.483 \pm 0.025$~\Rsun\ and $M_\star = 0.480 \pm 0.024$~\Msun.  These values agree quite closely with the values quoted by the \tess Input Catalog \citep[TIC,][]{Stassun19,TIC}.

\subsection{Stellar Rotation Period}

\subsubsection{TESS Photometry}
\label{ssec:rotation_tess}

The \tess photometry for this system is abundant, encompassing 27 sectors.  We examined the \tess Single Aperture Photometry (SAP) for signs of stellar rotation, as long-term trends can sometimes be attenuated or removed by Presearch Data Conditioning when it identifies and removes instrumental effects.  However, the SAP photometry may still contain meaningful instrumental systematics, so we will need to check to make sure our results are stellar in origin.

\begin{figure*}
    \centering
    \includegraphics[width=0.9\linewidth]{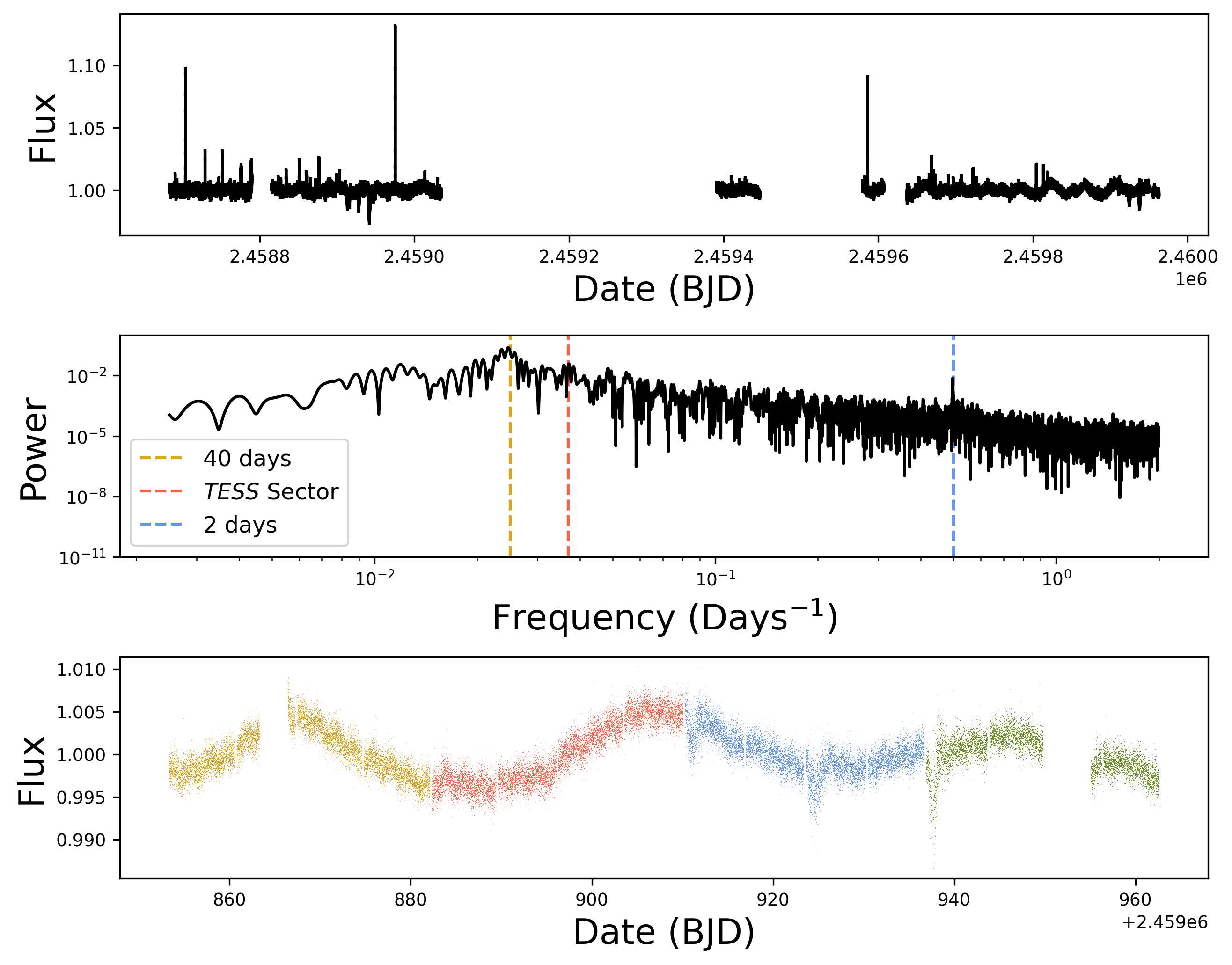}
    \caption{Top: The 120s \tess SAP fluxes for the TOI-1450 system for Sectors 14-60.  Each sector's data is mean-normalized. Middle: GLS periodogram of the \tess 120s data.  Dashed vertical lines are drawn on the periodogram corresponding with the peaks at 40 days, 27 days (the length of a single \tess sector), and 2 days.  Bottom: \tess photometry of sectors 57-60 the TOI-1450 system.  The different sectors are colored differently in order to draw attention to where potential breaks in the baseline are expected.  Sinusoidal variations on 40d and 2d timescales are obvious in these sectors.}
    \label{fig:tess_all}
\end{figure*}

The 120s cadence SAP photometric flux collected by \tess \citep[and read in using \texttt{lightkurve},][]{lightkurve} is shown in Figure~\ref{fig:tess_all}.  Each sector has been mean-normalized to emphasize the variability and periodicity of the data as opposed to the slight photometric offsets between sectors.  We examined the 120s data (which has a longer time baseline than the 20s data) for any evidence of rotational signals using the \texttt{PyAstronomy}\footnote{https://github.com/sczesla/PyAstronomy} \citep{PyAstronomy} implementation of the generalized Lomb-Scargle (GLS) periodogram \citep{Zechmeister09}.  The resulting periodogram is shown in the middle panel of Figure~\ref{fig:tess_all}.  

The strongest peak in the GLS periodogram corresponds to a signal of around 40\,days, which is visually obvious even when examining the raw data by-eye (see the bottom panel of Figure~\ref{fig:tess_all}).  There is also a signal at around 27\,days, which is likely merely related to the length of a \tess sector and is probably present due to systematics in the light curve.  Finally, there is a third signal corresponding to 2.01\,days, with a amplitude varying between 0.02\%---0.08\%.  This signal is very close to the observed planet transit period ($P\,=\,2.04$\,days), but appears to be distinct from the signal of the transiting planet, as it has a slightly different period (the 2.01\,day peak in the periodogram is very narrow and does not encompass the transiting planet's period) and there is obvious sinusoidal variation on this timescale in the photometric data.  Thus, this 2.01d signal may be related to a star's rotation.

We thus notice two apparent rotation signals in the photometry, one at 40\,days and one at 2.01\,days.  These could be representative of the rotation of the A and B stars, which are blended in the \tess pixel.  While the primary is thirty times brighter than the secondary, the secondary is younger and thus may possess a proportionally stronger time-varying rotational signal.  Thus, we cannot immediately determine which signal comes from which star based on the \tess data alone.  In general, we would expect that a shorter-period signal belongs to the smaller star \citep[see, e.g., the distributions of rotation periods with spectral type from][]{Popinchalk21}.  Taking the 97\% dilution from the A star into account, if the 2d signal does come from the B star, it actually corresponds to a 1---3\% amplitude signal in the \textit{TESS} flux band.  As a comparison, the rotation of TRAPPIST-1 (another late M dwarf) manifests in its \textit{TESS} photometry as a a roughly 0.6\% signal.  The slightly stronger amplitude of TOI-1450B may be the result of the star's shorter rotation period (and thus higher activity levels).  Therefore, it would be logical for the 2d signal to come from TOI-1450B, as its amplitude is reasonable compared to other short-period late M dwarfs.

Additionally, a 40\,day rotation signal would align with expectations from \cite{Popinchalk21} for a field M3 star.  However, the exact period of the 40\,day rotation signal is somewhat uncertain, as \tess has problems with detecting signals with periods longer than the length of the 27\,day \tess sector \citep{CantoMartins20}.  In Section~\ref{ssec:vsini}, we study the Doppler broadening of the MAROON-X spectrum to resolve which rotation period is associated with which star, finding that the smaller star is likely the faster rotator, in agreement with theoretical expectations.

It is obvious that the system has flared multiple times during the \tess observation window, with one flare energetic enough to produce 15\% of the combined system's luminosity.  As lower-mass stars tend to flare more frequently \citep{Davenport19} and these two stars are likely the same age, the majority of these flares are likely the result of TOI-1450B and not the exoplanet host star.  There is no clear photometric evidence that the star was flaring during any of our RV observations, so we were unable to confirm whether or not the flares come from TOI-1450A or TOI-1450B spectroscopically.

\subsubsection{ASAS-SN and WASP Photometry}

We studied both the ASAS-SN and WASP photometry in order to gain further insight on the potential 40d rotation period present in the \tess photometry.  Both instruments had pixel sizes large enough to result in blending between TOI-1450A and TOI-1450B.  The system appears stable, as the photometry, shown in Figure~\ref{fig:other_lc}, shows no evidence of of long-term trends on year-length timescales.  To search for signs of rotation, we produced GLS periodograms for each instrument, examining periods between 0.5---500\,days.  Overall, we expected both ASAS-SN and WASP to be more reliable than \tess over long periods of time. 

\begin{figure}
    \centering
    \includegraphics[width=1\linewidth]{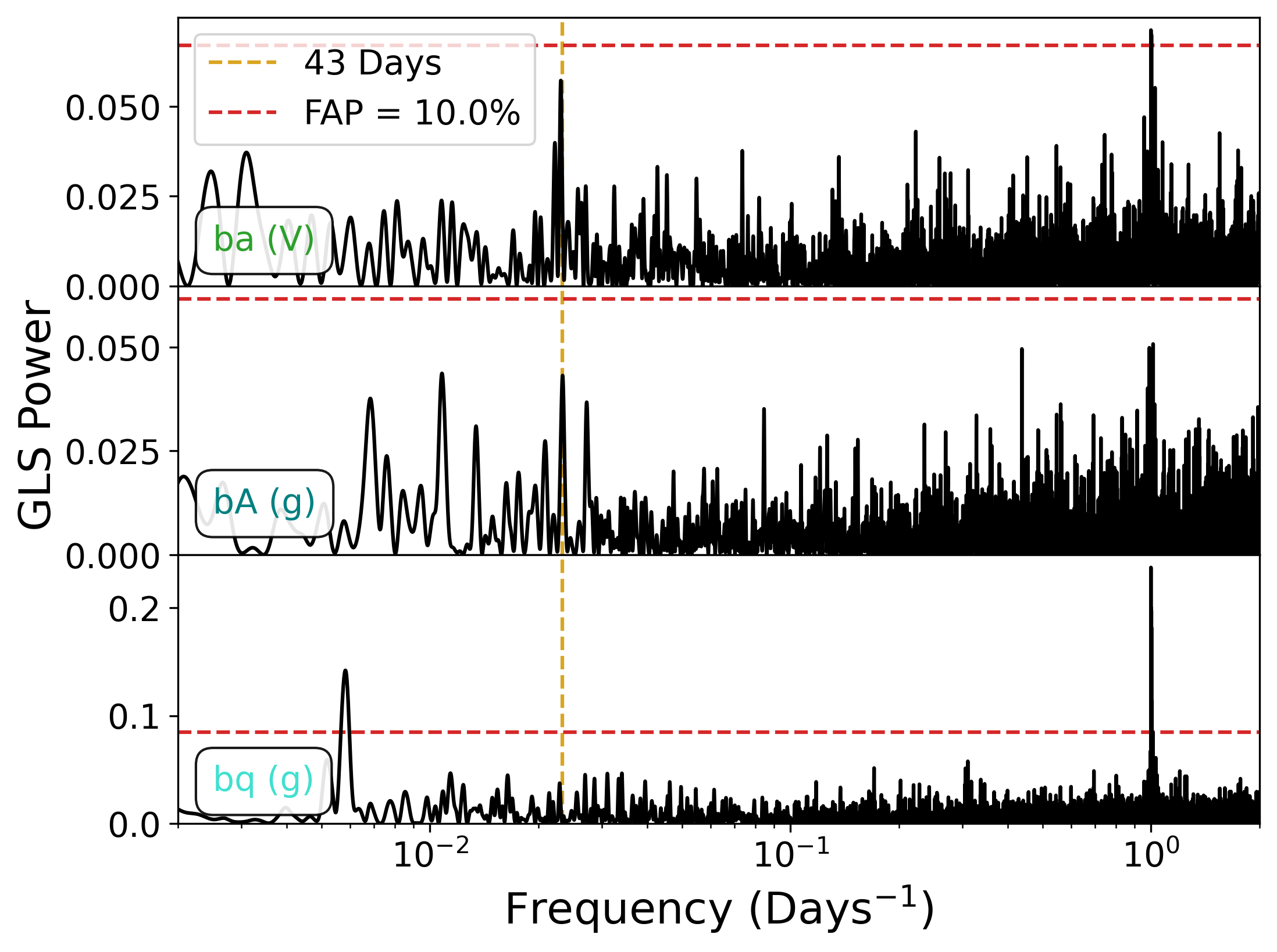}
    \caption{Lomb-Scargle periodograms of the ASAS-SN photometry.  The red dashed horizontal line corresponds to a 1\% false alarm probability, and the golden vertical line indicates a 43d period.}
    \label{fig:asas-sn_ls}
\end{figure}

Figure~\ref{fig:asas-sn_ls} shows the periodograms for three ASAS-SN cameras in two wavelength bands.  In general, we typically expect to see stronger activity-induced variability in the bluer $g$ band, and the $g$ band also has more statistical power due to the larger number of observations.  There are signals at around 1\,day and 172\,days that likely correspond with the nightly observing cadence and half of one year.  Additionally, both the ba and bA cameras find a high-FAP ($>\,10\%$) signal at $P\,\approx\,43$d, but the signal is not present in the bq data.  This signal coincides with the 40d signal observed by \textit{TESS}, strengthening the hypothesis that one of the stars in the TOI-1450 system has a 40d rotation period.  The data are noticeably missing the strong 2.01d signal noticed in the \tess photometry. However, this can be explained by the fact that the typical flux error for the ASAS-SN data ($\approx\,0.18\%$) is much larger than the amplitude of the rotational signal observed in the \tess data ($\approx\,0.03\%$).

\begin{figure}
    \centering
    \includegraphics[width=1\linewidth]{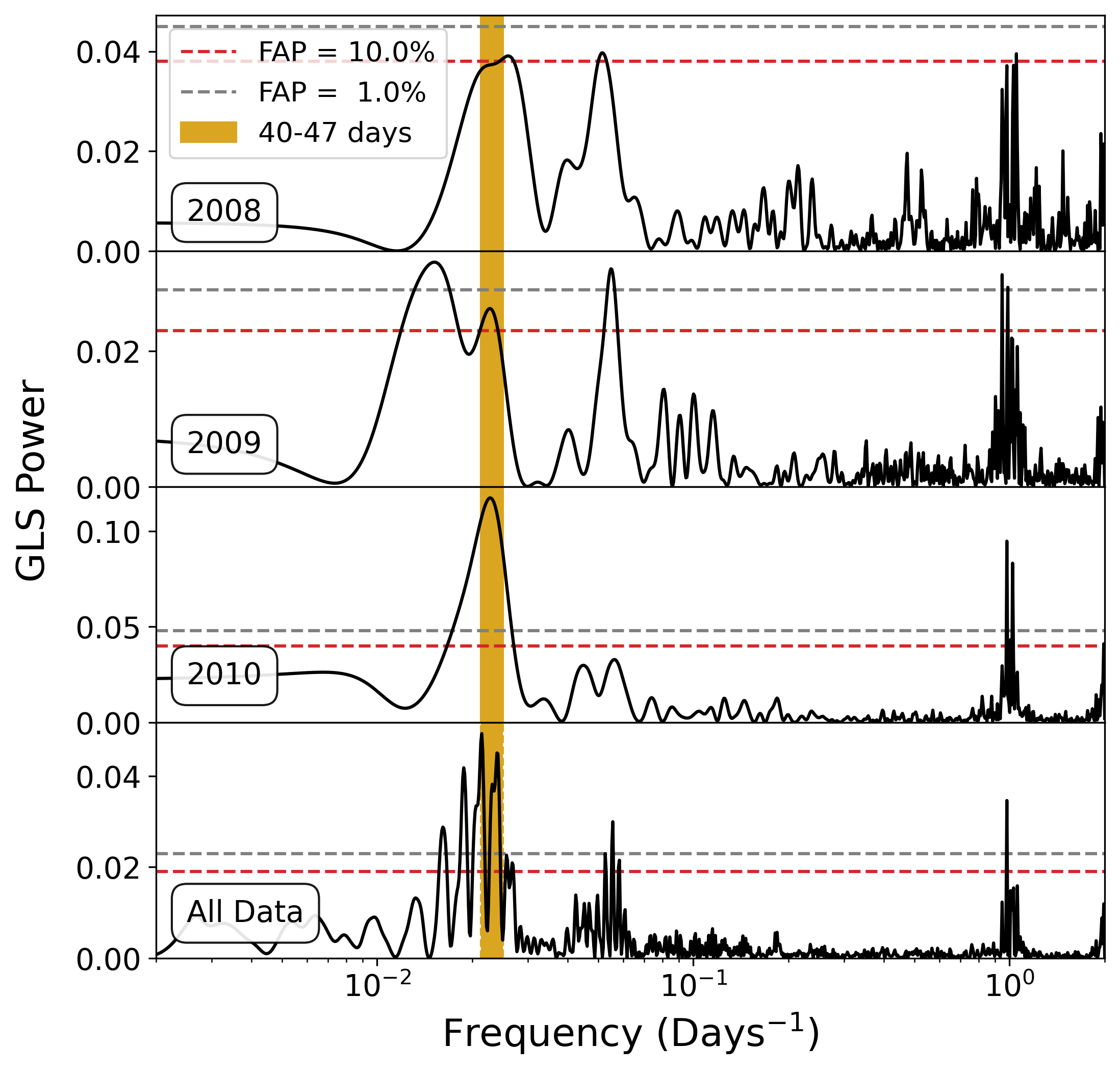}
    \caption{GLS periodograms of the WASP data.  The red dashed horizontal line corresponds to a 1\% false alarm probability, and the golden vertical span indicates periods from 40-47 days.  The signal at 1d is likely related to the nightly sampling and is thus unlikely to be physical.}
    \label{fig:wasp_periodogram}
\end{figure}

Figure~\ref{fig:wasp_periodogram} shows the periodogram of the WASP photometry.  Overall, we notice GLS periodogram peaks with a $<\,1$\% FAP in between $40\,$---$\,47$\,days in all three datasets, but the 40d peak is by far the strongest in the 2010 data.  Both the 2008 and 2009 datasets feature stronger signals at 20d, at half the proposed rotation period.  The 2009 data feature a strong peak at around 60\,days, but this signal is suspect given the fact that it is approximately half the length of the observing season.  Overall, the data are consistent with the $\approx$\,40\,day rotation period found by other instruments.

\subsubsection{Identifying the Fast Rotator by Estimating v sin i}
\label{ssec:vsini}

We can directly measure the projected rotational velocity $v\,$sin$\,i$ of TOI-1450A and TOI-1450B by examining the line broadening present in the MAROON-X spectra.  We can then check to see if this value is consistent with the observed 2d periodicity in the \tess data.  However, as MAROON-X has a resolution of roughly 85,000 \citep{Seifahrt18}, this method will be imprecise for spectra with $v\,$sin$\,i\,\lesssim\,2$\,\kms.   

To estimate $v\,$sin$\,i$, we used the cross correlation (CCF) comparison method described in \cite{Gray05} and used with MAROON-X data in previous works \citep[see Section 3.3 in][for a more comprehensive description]{Brady23}.  With this method, we compared the spectra of TOI-1450A and TOI-1450B to other spectra taken by MAROON-X of known slow rotators with similar spectral types, which allowed us to estimate the rotational velocities of TOI-1450A and TOI-1450B.  

As TOI-1450A is a M3 star, we compared it to Luyten’s Star, a M3.5V star \citep{Hawley96} with a rotation period of $115.9\,\pm\,19.4$ days \citep{Mascareno15}.  Using this star as a template, we find that the $v\,$sin$\,i$ of TOI-1450A is $1.8\,\pm\,0.6$\,\kms, which is below the $v$sin$i\,<\,2$\,km/s lower limit of recordable rotational velocities given the MAROON-X resolution limits \citep[see][for more details]{Brady23}.  It is thus possible that the $v$sin$i$ of TOI-1450A is below 2\,km/s, making the accuracy of this measurement suspect.  Given the radius of TOI-1450A (0.474\,\Rsun), a two-day rotation period would correspond to a measured rotational velocity of about 12 km/s. This would be easily detectable by MAROON-X if the planet system had a low obliquity.  The fact that the observed rotational velocity is below the measurement threshold of MAROON-X indicates that either the observed signal does not correspond to the rotational signal of TOI-1450A, or that the sin$\,i$ of TOI-1450A is extremely low ($i\,<\,10\degr$).  The chances of the system randomly being oriented at such a low inclination is fairly small ($<\,20\%$), but not impossible.  

We also used this method to estimate the rotational velocity of TOI-1450B and compared it to the observed photometric signal.  We used a spectrum of Teegarden’s Star as a template, as it is the latest-type slow-rotating M dwarf observed by MAROON-X \citep[with a rotation period of $99.6\,\pm\,1.4$ days][]{Terrien22}.  With it, we found that TOI-1450B has a rotation velocity of roughly $3.0\, \pm \, 0.7$ \kms.  Given TOI-1450B’s mass of about 0.14\,\Msun, we would expect it to have a radius of around 0.15\,---\,0.18\,\Rsun \cite[comparing the mass to those of other late M dwarfs, such as those in][]{Parsons18}, which would correspond to an equatorial velocity of around 3.8\,-\,4.5\,\kms if it had a two-day rotational period and we were observing it equator-on.  We cannot eliminate the possibility that TOI-1450B has a rotation period of approximately 4d and the observed 2d signal in the photometry is due to persistent features on opposite hemispheres, though this possibility is slightly more discrepant with our measured rotation $v$sin$i$.  Thus, the observed signal could easily be explained by the rotation of TOI-1450B, especially if it is somewhat inclined relative to the transiting planet.  Given its rapid rotation speed, it is unlikely to be the source of the 40d rotation signal.

From this analysis, we conclude that the short-period rotation signal in the photometry is likely to come from TOI-1450B, which makes sense given the fact that it is much smaller than TOI-1450A and is thus likely to rotate more rapidly.  This means that the 40d signal is likely the product of TOI-1450A's rotation.

\subsubsection{Activity}
\label{sssec:activity}

We can also estimate the rotation period of the star by studying the spectroscopic activity indicators.  \texttt{serval} measures several activity indicators of the input spectra, including the chromatic index, differential line width, $H\alpha$ index, the infrared Ca II triplet indices, and the Na$_I$ doublet indices.  We can examine periodograms of these activity indicators to get a sense of the stellar rotation period.  This examination is crucial to distinguishing genuine planetary signals from those that are merely stellar in nature, as we typically do not expect to see any planetary signals in the indicators.

\begin{figure*}
    \centering
    \includegraphics[width=0.7\linewidth]{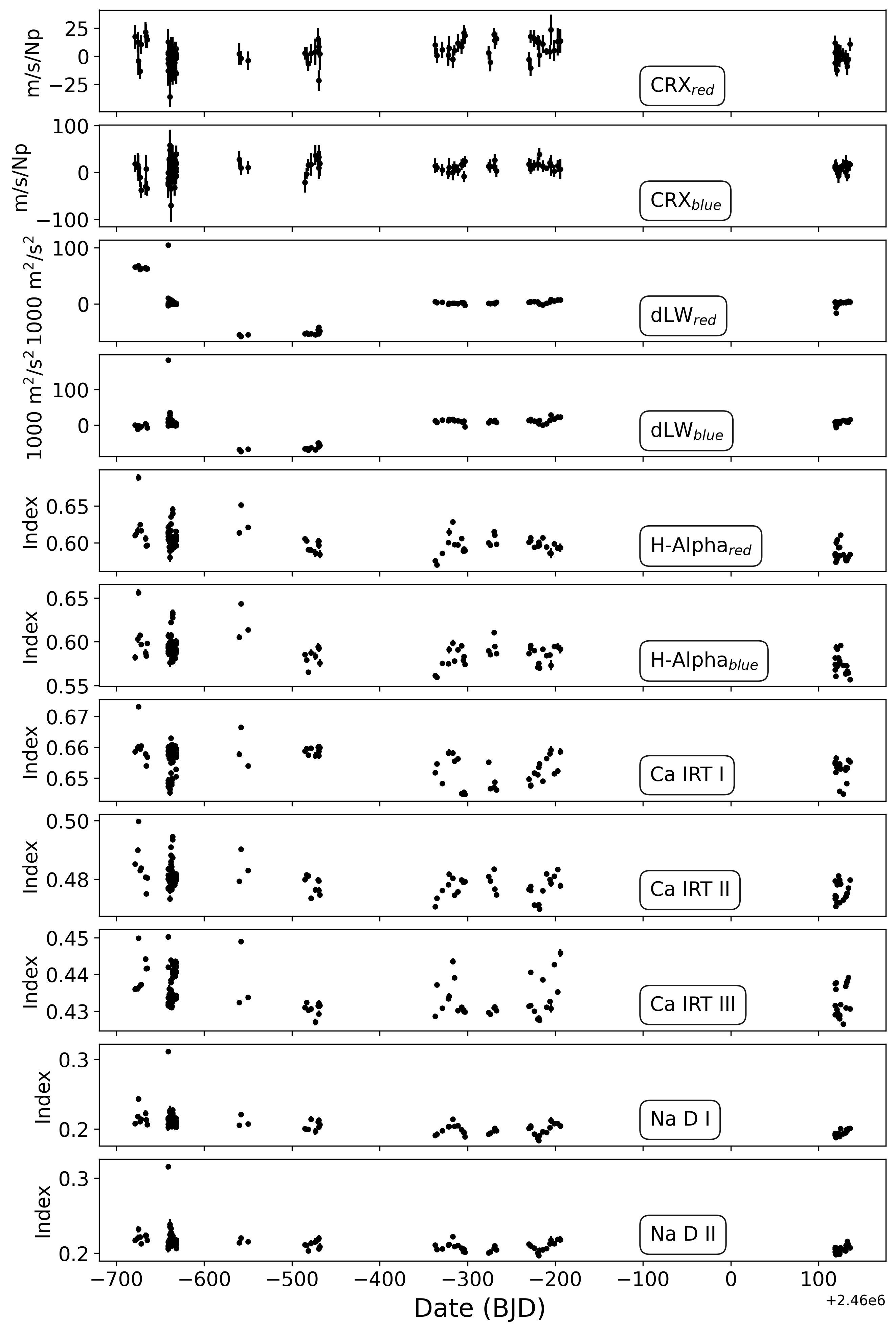}
    \caption{Time series of the MAROON-X activity indicators.}
    \label{fig:act_ind}
\end{figure*}

The actual values of the activity indicators are shown in Figure~\ref{fig:act_ind}. There are obvious offsets in the measured dLWs between each MAROON-X run.  This is a known issue and is believed to be due to small instrumental profile shifts as a result of the instrument front-end unit being unplugged from Gemini between each observation run.  The fact that the offsets are not the same in the red and blue channels between April and May 2021 could explain the fact that our RV offsets between runs are chromatic at that time.  There also appear to be long-period trends in the activity data that correspond to signals with $P\,=\,1$\,year, which are likely related to the barycentric motion of the Earth causing telluric lines to influence our activity measurements.

There is an obvious outlier in the May 2021 blue-channel data in both the dLW measurements and in the Na\,D indices.  A similar outlier of a smaller magnitude is also present in the red-channel dLW data at this time.  However, there is no obvious outlier in the corresponding RV measurement. An examination of the spectrum in question reveals strong emission in the wings of the sodium doublet at time of observation.  However, these emissive features are the result of sky emission of sodium and are masked out by \texttt{serval} when estimating RV.  They are also not included in \texttt{serval}'s calculation of the actual Na\,D index, and thus cannot explain the observed large index shift.  However, there are weaker telluric lines that cross the Na\,D line center during that exposure, and it is possible that they are responsible for the anomalous measurement.  However, as the shift is also recorded in the measured chromatic index, it is possible that this observation was actually taken during some active event, in which chromospheric Na\,D emission spiked.  We also note a small outlier in the H\,$\alpha$ index as recorded by both the red and blue channels in April 2021, though an examination of the involved spectra does not show any obvious flare or telluric activity.  As these observations may be the result of flares or failures by \texttt{serval} to accurately estimate the line indices, we removed them from our analysis for the purposes of creating the GLS periodograms.

\begin{figure*}
    \centering
    \includegraphics[width=0.7\linewidth]{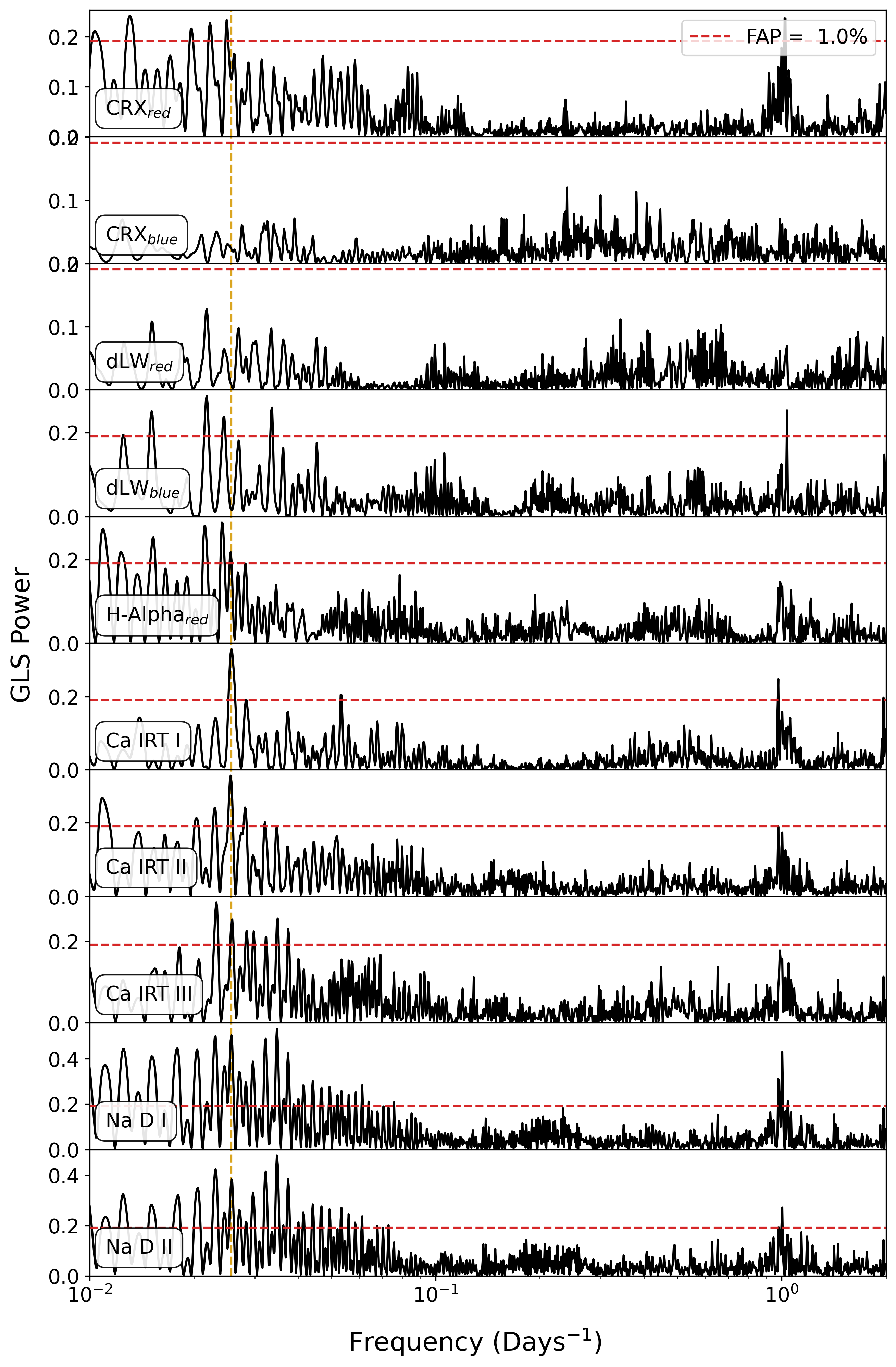}
    \caption{GLS periodograms of the signals present in the MAROON-X activity indicators, with outliers due to flares removed and the dLW data normalized to the mean within each run.  The dashed gold line indicates the location of a 39d period.  A red dashed line corresponding to a false alarm probability of 1\% is also shown.}
    \label{fig:act_per}
\end{figure*}

The GLS periodograms of the activity indices are shown in Figure~\ref{fig:act_per}.  In this figure, we removed the two observations that correlated with potential flare events, and also normalized the dLW data by subtracting the mean dLW value from each run.  We see that most of the activity indicators used by \texttt{serval} have signals at the FAP\,$<$\,1\% level at around 40 days, with the exact periods varying between about 39\,---\,47\,days.  While these periods are longer than the length of a typical ($\approx$30\,day) MAROON-X run, they are in agreement with the longer rotation period estimate from the \tess, ASAS-SN, and WASP data.  

A closer examination of the lines studied showed that the centers of three of our known activity indicator lines (the redder sodium doublet line and the bluest two Ca II lines) were contaminated by atmospheric tellurics during the majority of our observations, making their measured activity signals dubious.  As measured trends in line indices between the two sodium doublet lines agree despite the fact that one aligns with a known telluric region and the other does not, it appears that the telluric contamination in the line center is slight.  This telluric contamination does appear to be more of an issue with regards to the Ca II lines, however, as the three datasets are less obviously correlated.  To be conservative, if we ignore all activity indicators with known telluric crossings, we would only use the indices of H\,$\alpha$, the bluest sodium doublet line, and the reddest Ca II line.  All of these indicators have observed periodogram peaks around 39\,days, however, so dropping the other lines from our analysis does not affect our results.  It may be interesting in the future to attempt to fit out these lines \citep[using a package like \texttt{molecfit}, see][]{molecfitA, molecfitB} for the purposes of determining telluric-corrected activity indicators, but is unnecessary in this case due to the relatively unambigous nature of the signal in the other indices.

The activity indicators give us a clear sign that the rotation period of the A star is around 40 days.  This supports our theory that TOI-1450A is not the 2.01d rotator in the system, and that TOI-1450A has a rotation period of 40 days, which is in-line with what is expected for a field M3 dwarf \citep{Popinchalk21}.

If TOI-1450A has a rotation period of 40 days and TOI-1450B has a rotation period of 2 days, we can attempt to estimate the age of the system using gyrochronology.  Using the relations from \citep{Engle2023} for old M2.5-6.5 dwarfs, we recovered an age of $3.4\,\pm\,0.5$\,Gyr for TOI-1450A.  However, this relationship does not account for temperature.  We thus studied the gyrochrone of the 4\,Gyr-old M67 cluster from \cite{Dungee2022}, and find that TOI-1450A is a slightly faster rotator than the 3400\,K star observed in M67.  This agreed with the sub-4\,Gyr age estimate, indicating that TOI-1450A is a middle-aged star.  Unfortunately, it is much more difficult to provide a gyrochronological estimate of TOI-1450B's age, as \cite{Pass2022} shows that there is significant variability in the spindown rates of fully-convective M dwarfs.  While the 2\,d rotation rate is not inconsistent with a $\approx\,3$\,Gyr system age, acquiring a more precise system age from TOI-1450B is not possible.  However, the rotation rates of the two stars are not inconsistent with each other, suggesting an overall system age of a few Gyr.

\section{Planet Fitting}
\label{sec:fitting_planet}

\subsection{Transit Photometry}
\label{ssec:tr_ph}

We performed a fit to the transit photometry of the TOI-1450 system using \texttt{juliet} \citep{Espinoza19}, which makes use of transit modeling from \texttt{batman} \citep{batman} and \texttt{dynesty} \citep{Speagle2020} to perform nested sampling.  In our analysis of the \tess data, we considered the 2-minute cadence data for light curves taken prior to the \tess extended mission, and the 20-second data for those taken afterwards (using the highest-cadence data available for each sector).  We used the PDCSAP light curves in our analysis.  For both the 120s and 20s data, we performed 5$\sigma$ clipping to reduce the influence of outliers on our analysis of the transits. 

We searched the \tess data for planet signals using the \texttt{transitleastsquares} algorithm from \cite{Hippke19} to estimate the signal period and $t_0$. \texttt{transitleastsquares} searches for signs of planetary transits using a realistic model of transit ingress, egress, and stellar limb-darkening.  \texttt{transitleastsquares} found the 2.01d rotation signal at a high confidence, but when the period search range is limited to exclude this signal, it recovered the suspected planet at $P\,=\,2.04392$\,days and  $t_0\,=\,2458685.34341$\,BJD, which agrees with the values quoted online by the \tess team.

Before fitting the transits of the 2.04d planet, we performed light curve detrending.  As discussed in Section~\ref{ssec:rotation_tess}, there are nonplanetary signals in the data due to stellar rotation.  One of these signals, the rotation of TOI-1450B, has a period similar to the period of the suspected transiting planet.  Due to the large amount of photometric data available for the TOI-1450 system, we did not simultaneously perform gaussian process (GP) detrending and transit fitting of the data.  Instead, we detrended the \textit{TESS} data before fitting the transits.  We detrended the data by fitting a GP \citep[using an implementation of the approximate Matern kernel from \texttt{celerite}][]{celerite} to the 120s PDCSAP \textit{TESS} data from the system (the shorter-cadence 20s data is unnecessary for studying a ~2d sinusoidal rotation signal), assuming an uninformative prior on the amplitude and a normal prior on the period centered around the suspected 2d rotation period.  We also masked out the transits (assuming the parameters from \texttt{transitleastsquares}) when fitting for the GP trend.  The fit trend captured the ~2d variation signal in the photometry while preserving shorter-period variations (such as transits).

We then proceeded to fit the planet transits.  First, we detrended the full \tess dataset, including both 120s data and 20s data when it was available.  To do this, we divided each \tess data point by the GP model.  We did not perform the same extensive detrending for the ground-based transit data, as we do not have enough data for the ground-based instruments to estimate the GP amplitude of each instrument.  Once our detrending was complete, we identified \tess data points that were more than 45 minutes pre- or post-transit and omitted them from our transit analysis. This step was primarily performed as the extremely large amount of out-of-transit \tess data dramatically increased our computation time.  A phase-folding analysis of the transit data suggested a transit duration of 70\,minutes or less, so this masking was unlikely to remove any in-transit data.

We calculated the planet transit parameters using \texttt{juliet}, modeling the transit light curve of a planet with quadratic limb-darkening. The priors used for the fit are listed in Table~\ref{tab:priors_star}.  We fit the period $P$ and transit time $t_0$ using narrow priors informed by our \texttt{transitleastsquares} analysis, as the individual transits are very shallow and thus difficult to recover with uninformative priors.  

We fixed the eccentricity at $e\,=\,0$.  This is because it is difficult to constrain the eccentricity from transit data alone unless a secondary transit is detected, TTVs are detected, or the stellar density is highly constrained \citep[see, e.g.,][]{VanEylen15}.  We will revisit the validity of this assumption when analyzing the RV data in later sections.



The priors for the limb-darkening coefficients, $R_p/R_\star$ ratio, and impact parameter $b$ are set by the sampling methods used by \texttt{juliet}. To calculate the planet radius, we used the sampling technique described in \cite{Espinoza18}, which parameterizes $R_p/R_\star$ and $b$ using the uniformly distributed variables $r_1$ and $r_2$.  The limb-darkening coefficients were sampled according to \cite{Kipping13} using the uniformly distributed variables $q_1$ and $q_2$.  The stellar density $\rho_\star$ has priors based on the values for the mass and radius of TOI-1450A listed in Table~\ref{tab:host}.  As TOI-1450A is a member of the \tess Cool Dwarf list, these values come from the observationally-derived relations from \cite{Mann15} and \cite{Mann19} and are thus not likely to suffer from systematic modelling errors.

Our final analysis included the \tess data, the LCO-Sinistro data, and the LCO-MuSCAT3 data.  We also fit parameters to the mean flux and jitter of each instrument to capture any variations between instrument flux baselines.  We considered the 120s and 20s data to be two separate datasets for the sake of fitting jitter, mean, and limb-darkening terms.

\begin{center}
\begin{table}[]
    \begin{tabular}{|c|c|}
    \hline
    \textbf{Fit parameters} & \\
    \hline
    $P$ &  N(2.04392, 0.0001) \\
    $t_0$ (BJD) & N(2458685.34341, 0.005) \\
    $r_1$ &  U(0, 1) \\
    $r_2$ & U(0, 1) \\
    $\rho_\star$ (\gcc) & N(6.23, 0.62) \\ 
    $\mu$ & U(0, 1) \\
    ln$\sigma_{\mathrm{ppm}}$ & U(ln(1), ln(10000)) \\ 
    $q_{1}$ & U(0, 1) \\ 
    $q_{2}$ & U(0, 1) \\ 
    \hline
    \end{tabular}
    \caption{The priors used in the transit fit as well as the resulting fits.  The $\mu$, ln$\sigma$, $q_1$, and $q_2$ for the different instruments (the 120s \tess data, the 20s \tess data, the four separate LCO-MuSCAT3 colors and the two separate LCO-Sinistro transit observations) are being fit separately, but all share the same priors so we omit listing them individually in this table.}
    \label{tab:priors_star}
\end{table}
\end{center}

The resulting parameters for the transit fit are shown in Tables~\ref{tab:fit_transit_planet} and \ref{tab:fit_transit_instrument}.  The observed transit signal, in combination with the stellar radius from the TIC \citep[using]{Mann15}, is consistent with a $R_p \,=\, 1.13\, \pm \,0.04 R_\oplus$ planet (similar in size to Earth) with an elevated impact parameter ($b\, =\, 0.74 \,\pm \,0.02$).  Our fits are primarily dependent upon the \tess data, as the ground-based data didn't have enough statistical power to constrain the planet parameters.  The transit data alone did not constrain the stellar density, with our fits merely recovering the prior.  This is likely due to a combination of the short transit length and the shallow transit signal.

\begin{table}[]
    \centering
    \begin{tabular}{|c|c|}
    \hline
    \textbf{Fit parameters} & \\
    \hline
    $P$ (d)                 & $2.0439276 \pm 0.0000010$ \\
    $t_0$ (BJD)             & $2458685.34223 \pm 0.00042$ \\
    $r_1$                   & $0.8256\pm 0.0143$ \\
    $r_2$                   & $0.0219 \pm 0.0006 $ \\
    $\rho_\star$ (\gcc)     & $6.32 \pm 0.58$ \\ 
    \hline
    \textbf{Derived parameters} & \\
    \hline
    $a/R_\star$                 & $11.18 \pm 0.35$ \\
    $R_p$ ($R_\oplus$)          & $1.130 \pm 0.044$ \\
    $b$                         & $0.738 \pm 0.022$ \\
    $I$ (degrees)               & $86.213 \pm 0.227$ \\
    $T_\mathrm{dur}$ (h)               & $0.989 \pm 0.013$ \\ 
    \hline
    \end{tabular}
    \caption{The planet parameters from the \texttt{juliet} transit fit to the detrended \tess data.}
    \label{tab:fit_transit_planet}
\end{table}

\begin{center}
\begin{table*}[]

    \begin{tabular}{|l|c|c|c|c|c|c|}
    \hline
    Instrument     & $\mu$ (ppm)  & $\sigma$ (ppm)            & $q_{1}$                & $q_{2}$                & $u_{1}$         & $u_{2}$ \\
    \hline
    \tess$_{120s}$ & $-3 \pm 23$  & $6^{+17}_{-4}$     & $0.29^{+0.26}_{-0.18}$ & $0.32^{+0.37}_{-0.23}$ & $0.32 \pm 0.27$ & $0.16 \pm 0.31$ \\
    \tess$_{20s}$  & $10 \pm 18$  & $865^{+25}_{-25}$  & $0.32^{+0.19}_{-0.14}$ & $0.54^{+0.30}_{-0.34}$ & $0.57 \pm 0.31$ & $-0.04 \pm 0.33$ \\
Sinistro$_\mathrm{Jul}$& $31 \pm 96$& $1141^{+88}_{-83}$ & $0.57^{+0.29}_{-0.36}$ & $0.51^{+0.33}_{-0.34}$ & $0.66 \pm 0.48$ & $-0.01 \pm 0.42$ \\
Sinistro$_\mathrm{Aug}$& $6 \pm 86$ & $802^{+84}_{-83}$  & $0.54^{+0.31}_{-0.34}$ & $0.53^{+0.32}_{-0.34}$ & $0.66 \pm 0.47$ & $-0.04 \pm 0.41$ \\
    MuSCAT3$_g$    & $-4 \pm 61$  & $569^{+73}_{-72}$  & $0.56^{+0.30}_{-0.34}$ & $0.50^{+0.33}_{-0.32}$ & $0.64 \pm 0.46$ & $0.01 \pm 0.41$ \\
    MuSCAT3$_r$    & $39 \pm 66$  & $1718^{+56}_{-52}$ & $0.59^{+0.28}_{-0.34}$ & $0.51^{+0.33}_{-0.34}$ & $0.67 \pm 0.48$ & $-0.01 \pm 0.42$ \\
    MuSCAT3$_i$    & $25 \pm 71$  & $2248^{+55}_{-56}$ & $0.44^{+0.36}_{-0.30}$ & $0.44^{+0.35}_{-0.30}$ & $0.49 \pm 0.42$ & $0.06 \pm 0.37$ \\
    MuSCAT3$_z$    & $26 \pm 56$  & $2093^{+46}_{-45}$ & $0.45^{+0.34}_{-0.31}$ & $0.45^{+0.36}_{-0.31}$ & $0.51 \pm 0.42$ & $0.05 \pm 0.38$ \\
    \hline
    \end{tabular}
    \caption{The instrument parameters from \texttt{juliet} transit fit to the detrended \tess data.}
    \label{tab:fit_transit_instrument}

\end{table*}
\end{center}

We also tested whether or not it was feasible to detect TTVs in the TOI-1450 data using \texttt{exoplanet} \citep{exoplanet:exoplanet}.  We chose to use \texttt{exoplanet} (which uses MCMC sampling) instead of \texttt{juliet} (which uses nested sampling) at this step due to the large dimensionality of the problem.  We first generated a light curve with \texttt{batman} with noise taken directly from the out-of-transit light curve from \tess for TOI-1450 and added 15 minute TTVs, which are in line with some of the timing discrepancies observed by the ground-based observations.  Overall, we found that we could not accurately recover a 15 minute TTV signal from the simulated data, as the individual transits were too shallow compared to the noise to derive precise transit times.  Thus, we did not attempt to perform a more complete TTV analysis of the \tess data for this system.

\subsection{Radial Velocity}
\label{ssec:rv}

We fit the RV data for TOI-1450A using \texttt{juliet}.  The framework is very similar to what we used in Section~\ref{ssec:tr_ph}, but with different variables to account for the different observation methodologies.  The fit planet parameters include $K$, the planetary RV semi-amplitude, the period $P$, the time of conjunction $t_0$, the eccentricity $e$, and the argument of periastron $\omega$. We parameterized $e$ and $\omega$  as $\sqrt{e} $cos$ \omega$ and $\sqrt{e} $sin$ \omega$, and allowed these values to vary between -1 and 1.  The $t_0$ and $P$ priors were taken from the transit fits in Table~\ref{tab:fit_transit_planet}.  We also included some additional fit parameters to reflect instrumental effects.  As the red and blue channel cover different wavelength ranges, we allowed them to have separate fit mean and jitter terms.  We have priors on the RV offsets between several of the individual MAROON-X data runs (described in Section~\ref{ssec:maroonx}, see Table~\ref{tab:priors_joint} for the numerical values), so these priors were included on the fit means.  There were several runs in which offset calibration was impossible due to insufficient or ambiguous calibration data.  When no offset priors were available, we set broad uninformative priors on the individual offsets.

\subsubsection{Model Selection}
\label{ssec:rv_model}

\begin{figure*}
    \centering
    \includegraphics[width=0.7\linewidth]{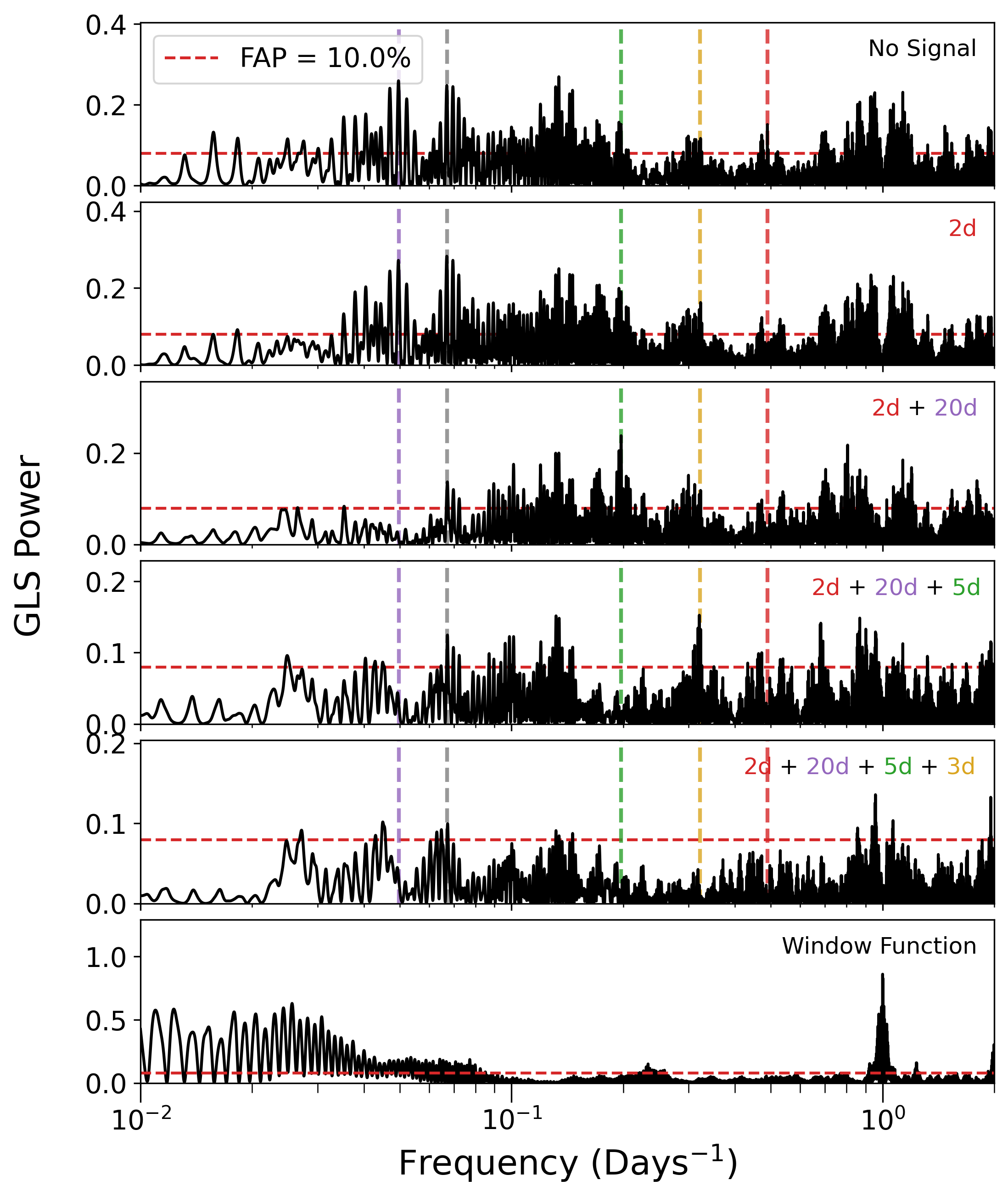}
    \caption{GLS periodograms of the RV data of TOI-1450 with various signals subtracted.  Each panel has a note describing which signals have been subtracted from the data to produce the residual periodogram.  The bottom panel shows the window function.  A red dashed line corresponding to a FAP of 0.1\% is shown on each plot.  In addition, vertical dotted lines corresponding to the periods of the fit planets are also included.  The red line is the 2.04d signal, the gold is the $\approx$3d signal, the green is the $\approx$5d, the grey is the $\approx$15d signal, and the purple is the 20d signal.}
    \label{fig:rv_gls}
\end{figure*}

To properly characterize the transiting planet, we must determine if there are any additional planets in the system and their characteristics.  We performed a simple period search of our RV data by progressively fitting circular Keplerian orbits to GLS periodogram peaks using \texttt{juliet}, searching for peaks between 0.5\,---\,50\,days.  Figure~\ref{fig:rv_gls} shows some of these periodograms, as well as our window function.  We ignored $P\approx1$d signals given the strong peak at that period in our window function. 

First, we performed a simple zero-planet fit including the offset priors to get a sense of the signals present in the data.  There was a peak at $P\,=\,2.04$d with a FAP\,$<$\,0.1\%, corresponding to the period of the known transiting planet.  The presence of this significant signal supports our statement that the transiting planet orbits this star and not TOI-1450B.   While this peak was not the dominant one in the periodogram, we were certain of its period and phase from the transit data and thus fit it out first, using priors on $P$ and $t_0$ from our analysis in Section~\ref{ssec:tr_ph}).  

After fitting out the transiting planet, the periodogram had several very low-FAP ($<\,0.1\%$) peaks remaining, one of which was at 20.2\,days (likely corresponding to one half of the rotation period) and the other at 14.9\,days. After attempting to fit it with a keplerian, we found that the 14.9d signal had a significantly higher amplitude in the blue channel of MAROON-X ($1.90\,\pm\,0.22$\,m/s) than it did in the red channel ($1.01\,\pm\,0.18$\,m/s).  This indicates that it may be heavily influenced by either stellar activity or some kind of instrumental systematic.  While it does not obviously have a timescale related to the stellar rotation period, it could potentially be an alias of the 20\,d signal related to the 57d peak in the window function (see the bottom panel of Figure~\ref{fig:rv_gls}).  The 20d signal is also chromatic ($1.64\,\pm\,0.18$\,m/s in the blue channel and $0.95\,\pm\,0.21$\,m/s in the red channel), supporting the hypothesis that it is related to stellar rotation.  However, this analysis is complicated by the fact that most MAROON-X runs are 1-4 weeks long, making it hard to disentangle the influence of a long-period activity signal on any long-period planets.  We thus hesitate to claim this signal as a planet despite its high significance.

As we are confident that the star has a rotation period of around 40d, we next performed a fit including the transiting planet and a keplerian at 20.2\,days (as a keplerian is an imperfect way to model an activity signal, we modeled it differently later on).  After this fit, the significance of the 14.9d signal was reduced and the highest residual peak was at 5.06d.  This 5.06d signal may correspond to a second planet in the system.  Fitting out the 5.06d signal (using a uniform prior from 4\,---\,5.5d) resulted in a forest of low-FAP peaks in the residual periodogram around 3.3\,days.  However, this final $\approx$\,3d signal was difficult to fit, with its period strongly depending on our treatment of the rotation signal.  It is possible that there is a planet around this period in the system (a planet at 3.18d or 3.39d would be in a $p,\,q\,=\,2,\,3$ or $p,\,q\,=\,1,\,2$ three-body Laplace resonance with the other two planets, respectively), but the signal is currently too low to confirm its presence.

To understand the influence of aliasing on our inferred model, we performed an analysis of the TOI-1450A system using the \lo periodogram\footnote{https://github.com/nathanchara/l1periodogram} \citep{Hara2017}.  The \lo periodogram is similar to a Lomb-Scargle periodogram in that it searches for signals in unevenly sampled data, but is specifically designed to reduce the number of observed peaks due to aliasing.  We performed a \lo periodogram analysis of all of the MAROON-X RV data, with the same offsets applied to the RVs as in Table~\ref{tab:priors_joint} and mean-subtractions when no calibrated offsets were available. 

For the purposes of noise modeling, considered a grid of models with different white noise ($\sigma_w\,=\,$0\,---\,1.5\,m/s, $\Delta\sigma_w\,=\,0.25$\,m/s), red noise ($\sigma_r\,=\,$0\,---\,1.5\,m/s, $\Delta\sigma_r\,=\,0.25$\,m/s), calibration noise ($\sigma_c\,=\,$0\,---\,1.5\,m/s, $\Delta\sigma_c\,=\,0.25$\,m/s), and red noise timescales ($\tau\,=\,$1, 2, 4, 10, 20, 40, 120).  For each noise model, we computed the cross-validation score using the same methods as described in \cite{Hara2020}.  In general, for each noise model, we selected signals with log$_{10}$FAP$\,<\,-0.5$.  Then, the data were split into a randomly-selected training and test set (with a 60/40 split).  A sinusoidal model with the selected frequencies was fit on the training set, and the likelihood of this model was evaluated on the test set. This process was repeated 400 times for each noise model, and we quote the median of the likelihoods for each noise model as the cross validation score.

\begin{figure}
    \centering
    \includegraphics[width=0.9\linewidth]{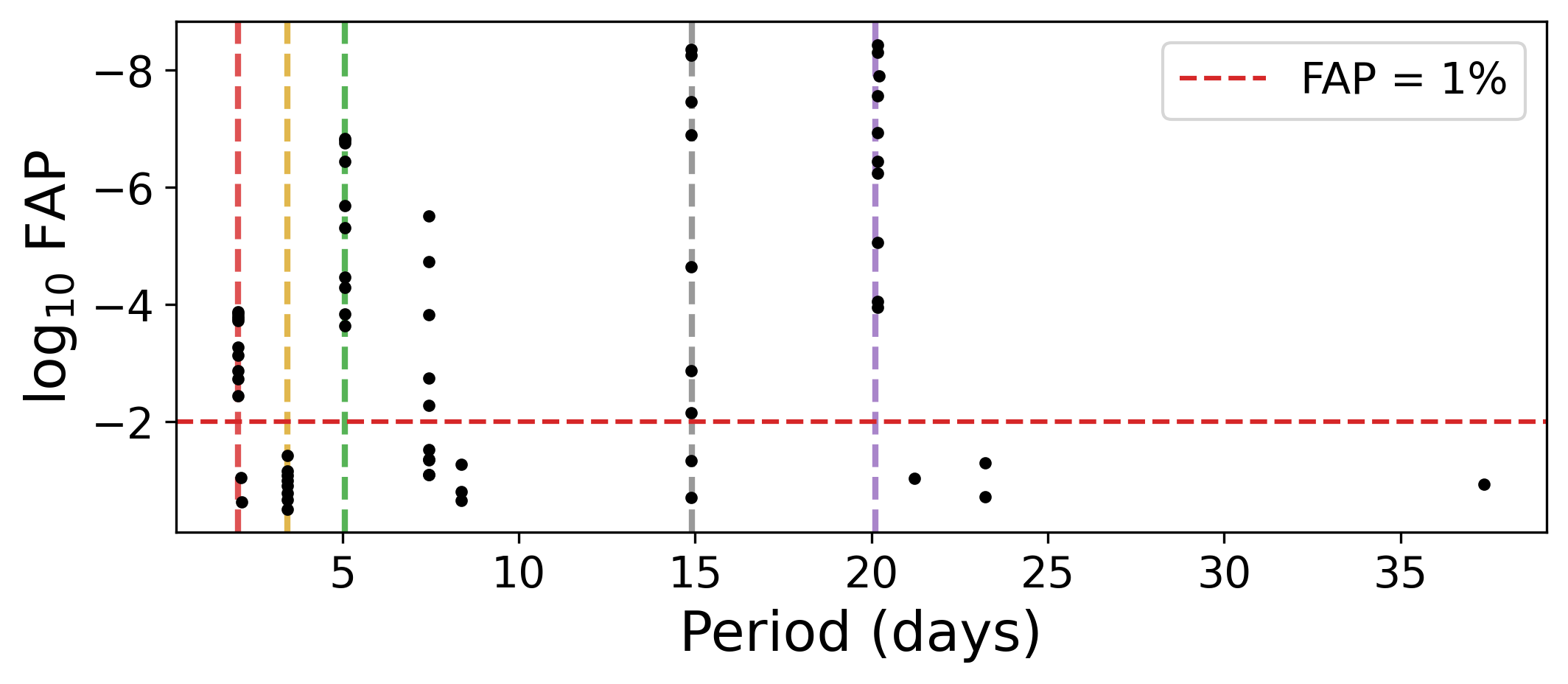}
    \caption{A \lo periodogram of the TOI-1450 RV data.
     Vertical dotted lines corresponding to various periods.  The red line is a 2.044d signal, the gold a 3.44d signal,  the green a 5.06d signal, the grey a 14.9d signal, and the purple a 20.1d signal.  The horizontal dashed line corresponds to a 1\% FAP.}
    \label{fig:l1_periodogram}
\end{figure}

Figure~\ref{fig:l1_periodogram} shows the resulting FAPs and periods of the selected signals for the top ten highest-ranking models (representative of about 0.5\% of the models evaluated).  We note the presence of both the transiting planet signal and the 5.06d planet signal at FAPs that are less than 0.1\%, indicating that these planets are present in the data.  We also see signals at 15d and 20d that correspond with the chromatic signals we have observed in our previous analysis that are unlikely to be planetary in origin.  We also see a low-FAP signal at around 7.5d, which corresponds with half of the 15d signal and is thus also unlikely to be planetary in origin.  Interestingly, we do see the presence of a signal at $P\,=\,3.44$d in this analysis, but the FAP is greater than 1\% and thus we hesitate to claim it as a planet.  More data may be necessary to confirm or deny the presence of this signal.  Several other high-FAP signals were also identified in the \lo periodogram, but we do not claim planetary detections at these periods for similar reasons.

In summary, we identified the presence of signals corresponding to the known transiting planet ($P\,=\,2.04$d), as well as signals at approximately 5, 15, 20, and 3 days, though we suspect that the 15d, 20d, and 3d signals may be related to stellar activity and/or systematics (or at least contaminated heavily by these effects).  To investigate the significance of these signals, we performed nested sampling fits with normal priors centered around the identified periodogram peaks, and found the log evidence of each model with \texttt{juliet}.  

We also included models in which we used Gaussian Processes (GPs) to model the stellar activity using a quasi-periodic kernel \citep[see, e.g.,][]{Aigrain2012, Stock2023}.  The quasi-periodic kernel is frequently used to model the effect of stellar activity on RVs and photometry, as it is able to reproduce the periodic signature of stellar rotation.  The kernel is

\begin{equation}
\kappa (t-t') = \sigma_{i}^2 \, \mathrm{exp} \Bigg(-\frac{\alpha(t-t')^2}{2} - \Gamma \mathrm{sin}^2 \bigg(\frac{\pi (t-t')}{P_\mathrm{rot}}\bigg) \Bigg),
\end{equation}

where $\alpha$ is the inverse square of the correlation timescale (which usually corresponds with the spot evolution timescale), $P_\mathrm{rot}$ is the rotation period of the star, $\Gamma$ describes the relative strength of the periodic to the aperiodic portion of the modeled signal, and $\sigma$ is a wavelength-dependent parameter that describes the amplitude of the signal.  The first term of the equation describes the aperiodic portion of the signal and the second term describes the periodic portion.  We fitted $\sigma$ separately for the red and blue channels of MAROON-X but assumed all of the other GP parameters were instrument-independent. 

We set our GP priors by following the suggestions of \cite{Stock2023}.  We allowed $P_\mathrm{rot}$ to have a normal prior of $40\pm3$\,days, which both encompasses the range of periods suggested by both the activity indicators and photometry.  We set $\alpha$ such that the correlation timescale falls between 40 days and the length of our data collection, as it has been shown observationally that M dwarfs typically have spot evolution timescales longer than one rotation period \citep[see, e.g.,][and the references therein]{Robertson2020}.  We have allowed $\Gamma$ to have a log-uniform prior from 0.01 to 10 following the suggestions of \cite{Stock2023}, as very high values of $\Gamma$ can result in GP overfitting.  We have placed an upper limit of 4\,m/s on $\sigma$, reflecting the degree of scatter observed in the RVs.  The priors for our GPs are listed in Table~\ref{tab:priors_joint} in Section~\ref{ssec:joint}.

The models examined, with their associated transiting planet masses and relative Bayesian evidences, are shown in Table~\ref{tab:rv_models}.  The model with the highest Bayesian evidence is the model that the data supports the most heavily, though we note that differences in ln$Z$ on the order of 1\,---\,2.5 only indicate weak to moderate evidence for a model, as discussed in \cite{Trotta2008}.

\begin{table*}[]
\centering
\begin{tabular}{|l|c|c|c|}
\hline
Model                    & $M_\mathrm{b}$sin$i_\mathrm{b}$ ($M_\oplus$) & dln$Z$ & P$_\mathrm{rot}$        \\ \hline
-                        &                             & -78.1   &                                               \\
2d                       & $1.24 \pm 0.19$             & -60.3   &                                               \\
2d, 5d                   & $1.30 \pm 0.16$             & -43.0   &                                               \\
2d, 15d                  & $1.18 \pm 0.15$             & -27.8   &                                               \\
2d, 20d                  & $1.01 \pm 0.19$             & -41.1   &                                               \\
2d, 3d, 5d               & $1.29 \pm 0.13$             & -40.6   &                                               \\
2d, 5d, 15d              & $1.34 \pm 0.09$             & -25.3   &                                               \\
2d, 5d, 20d              & $1.16 \pm 0.15$             & -16.4   &                                               \\
2d, 3d, 5d, 15d          & $1.27 \pm 0.09$             & -37.6   &                                               \\ 
2d, 3d, 5d, 20d          & $1.25 \pm 0.13$             & -11.5   &                                               \\ 
2d, 3d, 5d, 15d, 20d     & $1.27 \pm 0.16$             & -7.6    &                                               \\ 
\hline

GP                       &                             & -37.5   & $39.9^{+1.7}_{-1.4}$                          \\
2d + GP                  & $1.08 \pm 0.13$             & -13.5   & $38.0^{+1.4}_{-1.4}$                          \\
\textbf{2d, 5d + GP}     & $1.25 \pm 0.13$             &         & $40.0^{+1.7}_{-1.4}$                          \\
2d, 15d + GP             & $1.05 \pm 0.13$             & -13.5   & $38.0^{+1.7}_{-1.6}$                          \\
2d, 20d + GP             & $1.15 \pm 0.12$             & -7.0    & $29.9^{+2.4}_{-1.4}$                          \\
2d, 3d, 5d + GP          & $1.26 \pm 0.10$             & 13.8    & $38.9^{+0.5}_{-0.5}$                          \\
2d, 5d, 15d + GP         & $1.23 \pm 0.12$             & 8.5     & $39.9^{+1.8}_{-1.5}$                          \\
2d, 5d, 20d + GP         & $1.24 \pm 0.13$             & 10.0    & $39.5^{+1.9}_{-1.8}$                          \\
\hline

\end{tabular}
\caption{List of all of the models used to fit the data with \texttt{juliet}, with the associated log evidences and mass of the transiting planet signal.  The log evidences are listed in terms of the difference between the final chosen model's evidence and the individual models' evidences, so lower numbers indicate more preferred models.  The final chosen model is bolded.  The typical error on ln$Z$ is 0.4\,---\,0.8, and is omitted from the table to save space.}
\label{tab:rv_models}
\end{table*}

The four models with the highest evidence are a two-planet model featuring the transiting planet and the 5d planet, as well as three three-planet models with the transiting planet, 5d planet, and a 3d, 15d, or 20d planet in addition.  However, we are hesitant to select any of these three-planet models, as we have found in our previous analysis that the 15d and 20d signals in our data are highly chromatic and the 3d signal has a broad period prior and cannot be recovered by the \lo periodogram with a <1\% FAP.  A closer look at our 3d planet fit shows that the observed $K$ of the 3d planet is degenerate with the GP $\sigma$, meaning that the fit signal may be spurious and dependent upon how we chose to fit the stellar rotation.  While it is possible that there is a 3d planet in our data, we are hesitant to claim its presence given our data.  If it does exist, it has a $K\,\leq\,0.6$\,m\,s$^{-1}$ and would thus likely need far more data to characterize accurately.

We also studied the influence of allowing the eccentricites of the planets to vary on our selected 2-planet model.  We found that, when allowing $\sqrt{e} $cos$ \omega$ and $\sqrt{e} $sin$ \omega$ to vary uniformly between -1 and 1 with our selected model, there was an increase in the BIC ($\Delta$BIC\,=\,7.1), but the fit planet eccentricities were low (with broad posterior distributions) and the planet masses were consistent (within 0.2$\sigma$) with our circular-orbit model.  We also found that allowing for nonzero eccentricities in our model without the fit GP resulted in essentially no change ($\Delta$BIC\,=\,0.8).  The fact that the planets only appear to have eccentric orbits with our GPs in place indicates that the small eccentricities observed may be the result of cross-talk between our keplerian and stellar activity model and may not be physical.  We thus conclude that we do not have sufficient evidence to conclude the planets have nonzero eccentricities.  This aligns with the expectation that these short-period planets would likely have circular orbits, as the 2d planet has a tidal circularization timescale of $<\,2$\,Myr and the 5d planet (assuming a rocky planet with a a nearly edge-on orbit and a larger radius than the 2d planet) has a circularization timescale of $<\,100$\,Myr.  Given that this system is likely several Gyr old, we expect the orbits to be near-circular.

Thus, we select the two-planet model as our final model, as the models with higher evidences are likely the result of overly flexible GP fitting.  We note that all of the models with higher BIC values have fit transiting planet masses within 1$\sigma$ of our chosen model, so the transiting planet mass is not sensitive to our choice of model.   
Overall, the data support the presence of a transiting planet, as well as a planet with a 5.1d period.  There are longer-period signals present in the data, but it is difficult to disentangle aliases of potential rotational signals from long-period planets, as they seem to encompass similar ranges of periods.  There also appears to be tentative evidence for a planet with a roughly 3d period in the data (which could be in period commensurability with the two known planets), but we currently lack the statistical evidence to confirm it.

\subsection{Joint Transit and Radial Velocity Models}
\label{ssec:joint}

In this section, we perform a joint fit of the RV and transit data using \texttt{juliet} to determine our final planet parameters. 

Before performing a joint fit to the RV and transit data, we checked the data with \texttt{transitleastsquares} to see if there were any transit signals consistent with the 5d signal.  A closer examination of the \tess light curves yields no obvious signals of a transiting planet at the newly-identified planet's period and $t_0$.  This is not surprising given the high impact parameter of TOI-1450Ab.  If TOI-1450Ab and TOI-1450Ac were coplanar, TOI-1450Ac would have an impact parameter $b_c\,>\,1$ and thus not transit.

As we do not find any evidence of any additional transiting planets in the \tess data, we used the RV data alone to characterize the non-transiting planet in the system. For the priors on the RV fit, we adopted the model from Section~\ref{ssec:rv_model}.  This model included the transiting planet, a 5d planet, and a quasi-periodic GP to model the $\approx$40d rotational signal.  The orbital eccentricities were fixed at 0.  We use the same priors for the transit model that we used in Section~\ref{ssec:tr_ph}, with the same data and detrending techniques.  Our priors for the joint fit are listed in Table~\ref{tab:priors_joint}.

\begin{table*}[]
    \centering
    \begin{tabular}{|c|c|}
    \hline
    \textbf{Fit parameters} &  \\
    \hline
    $P_b$ (d)                    &  N(2.04392, 0.0001) \\
    $P_c$ (d)                    &  U(4, 5.5) \\
    $t_{0, b}$ (BJD)             & N(2458685.3434, 0.005) \\
    $t_{0, c}$ (BJD)             & U(2459321, 2459326.5) \\
    $K_b$ (m\,s$^{-1}$)          & U(0, 3) \\
    $K_c$ (m\,s$^{-1}$)          & U(0, 3) \\
    $e_{bc}$                     & 0 (fixed)  \\
    $\omega_{bc}$                & 90 (fixed) \\
    $\rho_\star$ (\gcc)          & N(6.4, 1) \\ 
    
    \hline
    $\mu_{\mathrm{Apr 21, R}}$ (m\,s$^{-1}$)                                       & U(-10, 10)          \\
    $\mu_{\mathrm{Apr 21, B}}$(m\,s$^{-1}$)                                        & U(-10, 10)          \\
    $\mu_{\mathrm{May 21, R}} - \mu_{\mathrm{Apr 21, R}}$(m\,s$^{-1}$)             & N(2.5, 1)          \\
    $\mu_{\mathrm{May 21, B}} - \mu_{\mathrm{Apr 21, B}}$(m\,s$^{-1}$)             & N(2.5, 1)          \\
    $\mu_{\mathrm{Aug 21, R}}$(m\,s$^{-1}$)                                        & U(-10, 10)          \\
    $\mu_{\mathrm{Aug 21, B}}$(m\,s$^{-1}$)                                        & U(-10, 10)          \\
    $\mu_{\mathrm{Nov 21, R}} - \mu_{\mathrm{Aug 21, R}}$(m\,s$^{-1}$)             & N(2.5, 1)          \\
    $\mu_{\mathrm{Nov 21, B}} - \mu_{\mathrm{Aug 21, B}}$(m\,s$^{-1}$)             & N(1.5, 1)          \\
    $\mu_{\mathrm{Mar 22, R}}$(m\,s$^{-1}$)                                        & U(-10, 10)          \\
    $\mu_{\mathrm{Mar 22, B}}$(m\,s$^{-1}$)                                        & U(-10, 10)         \\
    $\mu_{\mathrm{May 22, R}} - \mu_{\mathrm{Mar 22, R}}$(m\,s$^{-1}$)             & N(1.5, 1)          \\
    $\mu_{\mathrm{May 22, B}} - \mu_{\mathrm{Mar 22, B}}$(m\,s$^{-1}$)             & N(-1.5, 1)          \\
    $\mu_{\mathrm{Jul/Aug 22, R} - \mu_{\mathrm{Mar 22, R}}}$(m\,s$^{-1}$)         & N(3.7, 1.12)          \\
    $\mu_{\mathrm{Jul/Aug 22, B}} - \mu_{\mathrm{Mar 22, B}}$(m\,s$^{-1}$)         & N(0.5, 1.41)          \\
    $\mu_{\mathrm{Jul 23, R}}$(m\,s$^{-1}$)                                        & U(-10, 10)          \\
    $\mu_{\mathrm{Jul 23, B}}$(m\,s$^{-1}$)                                        & U(-10, 10)         \\
    ln$\sigma_{\mathrm{all}}$(m\,s$^{-1}$)                                         & U(ln(0.001), ln(5))       \\
    \hline
    ln$\sigma_{GP,~red}$  & U(ln(0.01, ln(4))\\
    ln$\sigma_{GP,~blue}$ & U(ln(0.01, ln(4))\\
    ln$\alpha_{GP}$ & U(ln($1.5\times10^{-6}$), ln($6.25\times10^{-4}$)) \\
    ln$\Gamma_{GP}$ & U(ln(0.01), ln(10)) \\
    $P_\mathrm{rot}$ & N(40, 3) \\
    \hline
    \end{tabular}
    \caption{The priors used in the RV fit.  The joint fit model included both these priors and the transit priors from Table~\ref{tab:priors_star}.  The ln$\sigma$ for the  different MAROON-X runs and channels are being fit separately, but all share the same priors so we omit listing them individually in this table.}
    \label{tab:priors_joint}
\end{table*}

\begin{figure*}
    \centering
    \includegraphics[width=0.9\linewidth]{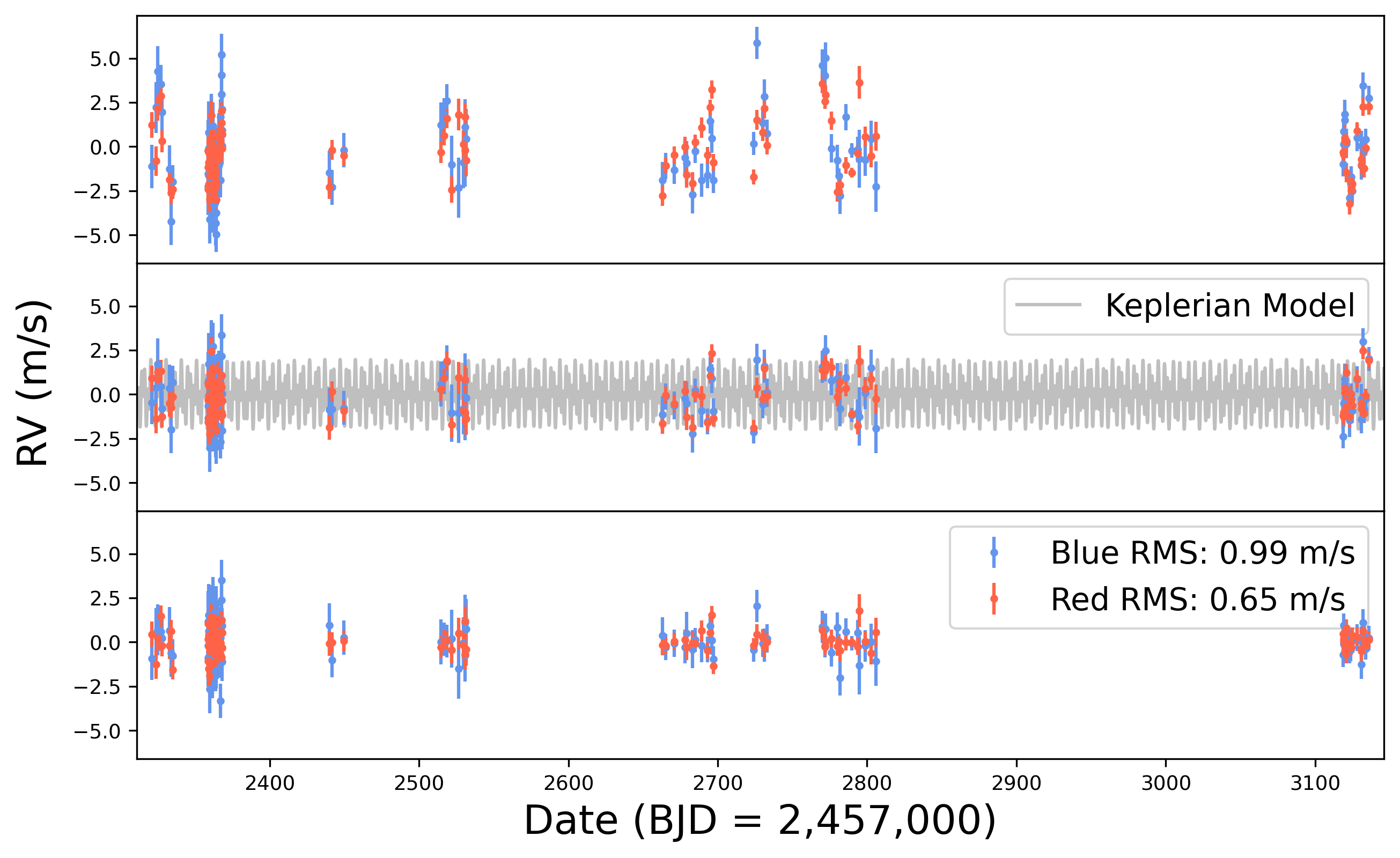}
    \caption{Top: The offset-subtracted RV data for TOI-1450A.  The MAROON-X red and blue channels are colored accordingly.  Middle: The RV data with both the offsets and best-fit GPs subtracted.  The best fitting 2-planet Keplerian model is shown in gray.  Bottom: The data residuals (subtracting out the offsets, GPs, and best-fit Keplerians).}
    \label{fig:rv_all}
\end{figure*}

\begin{figure}
    \centering
    \includegraphics[width=0.9\linewidth]{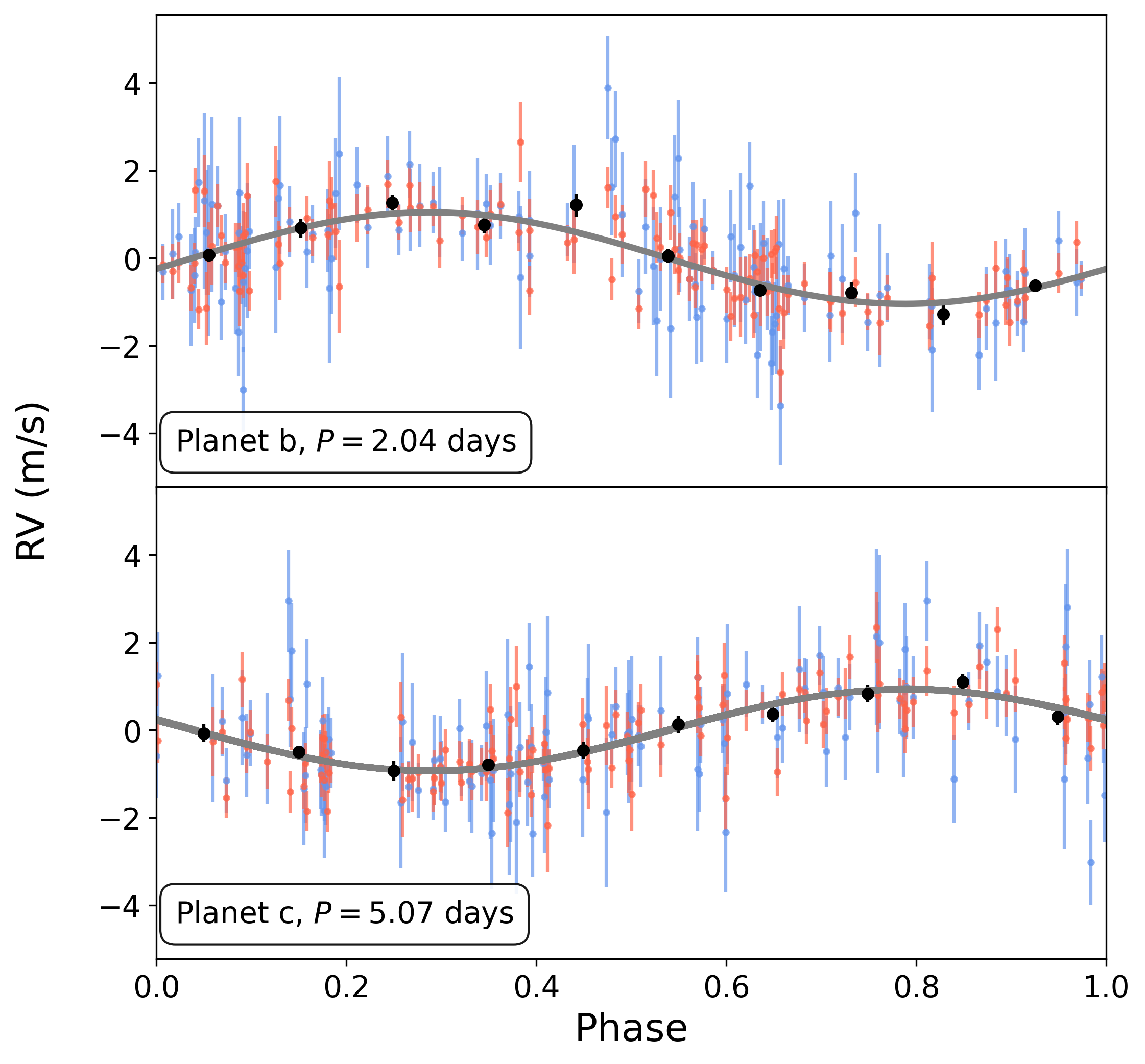}
    \caption{The phase-folded RVs for TOI-1450Ab and TOI-1450Ac.  The offsets, activity, and other planets are subtracted out in each panel to focus on the relevant signal.  The MAROON-X red and blue channels are colored accordingly, and binned data is shown in black.  The best-fit model is shown in gray.}
    \label{fig:rv_phase}
\end{figure}

\begin{figure*}
    \centering
    \includegraphics[width=0.8\linewidth]{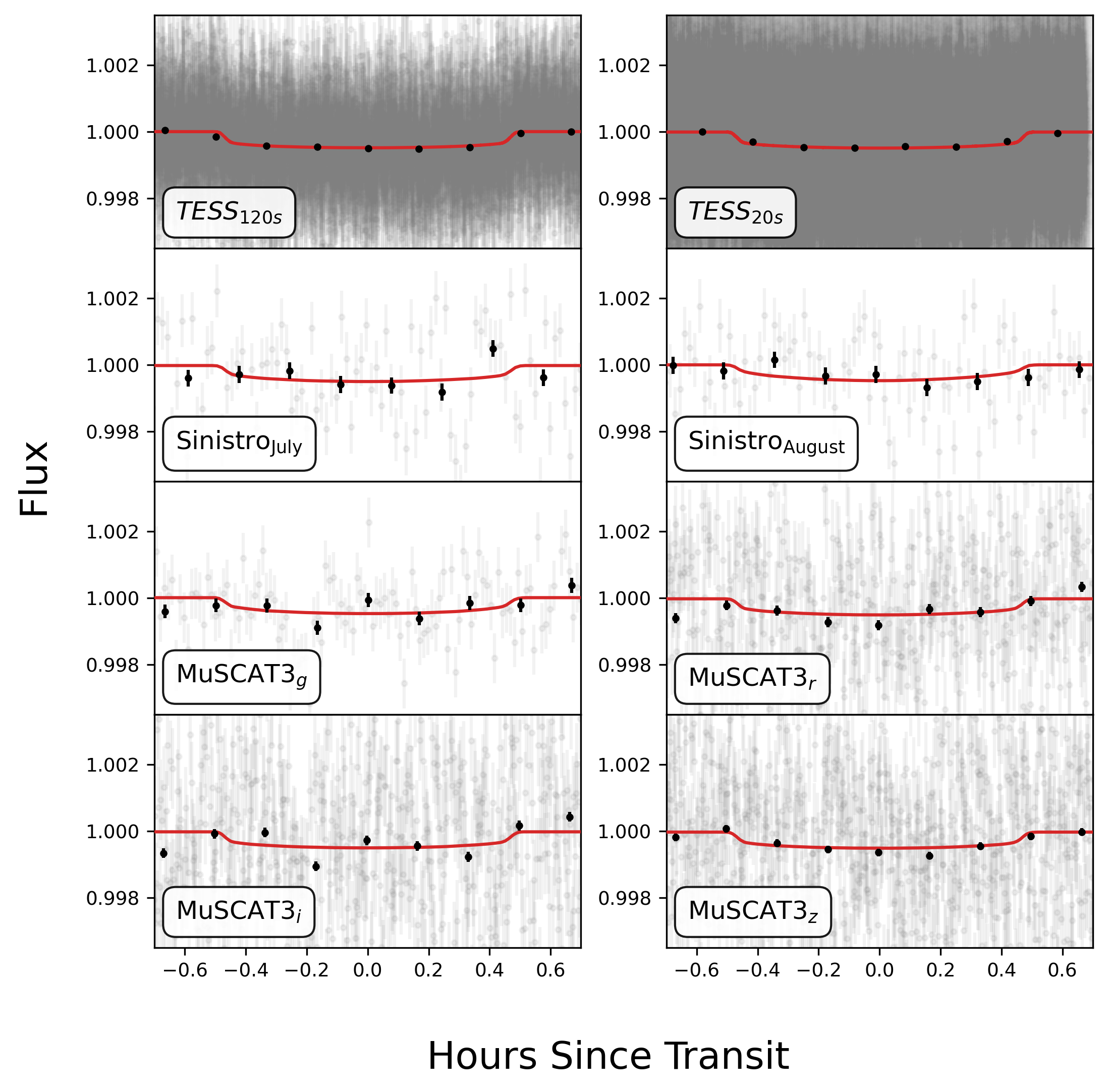}
    \caption{The phase-folded transit data for TOI-1450Ab, with our best-fit model shown in red.  Binned data are shown as black points. Each panel highlights a different dataset.}
    \label{fig:lc_phase}
\end{figure*}

The numerical results of our joint fits are listed in Tables~\ref{tab:fit_joint_planet} and \ref{tab:fit_joint_instrument}, and the resulting plots are shown in Figures~\ref{fig:rv_all}, \ref{fig:rv_phase}, and \ref{fig:lc_phase}.  We found that, using the TIC-derived stellar parameters, TOI-1450Ab has a mass of $1.26\,\pm\,0.13$\,\MEarth and a radius of $1.13\,\pm\,0.04$\,\REarth.  This corresponds to a density of $4.8\,\pm\,0.7$\,\gcc.  If we instead use the SED-derived stellar radius and mass, we recover a planet mass of $1.27\,\pm\,0.13\,M_\oplus$ and radius of $1.16\pm0.07\,R_\oplus$, leading to a density of $4.6\,\pm\,1.0$\,\gcc.  These values are all within 1$\sigma$ of the planet parameters using the TIC-derived stellar parameters. This demonstrates that our measured planet mass is somewhat robust to changes in how the stellar parameters are calculated.  For the purposes of this paper, we will use the planet parameters derived using the TIC stellar parameters given their higher precision and empirically-calculated nature. 

Our extensive dataset allows us to measure the mass of the transiting planet with a 10\% precision despite its very small size and the complicating influence of the additional planet, activity signal, and near-integer day orbital period.  We are confident in our precise mass measurement, as we found that our derived planet mass was insensitive to our model choice or method for determining stellar mass. 

\begin{table*}[]
    \centering
    \begin{tabular}{|c|c|c|}
    \hline
    \textbf{Fit parameters} & {b} & {c} \\
    \hline
    $P$ (d)                 & $2.0439274 \pm 0.0000010 $ & $5.0688 \pm 0.0019$\\
    $t_0$ (BJD)             & $2458685.34221 \pm 0.00042$ & $2459321.76 \pm 0.16$\\
    $r_1$                   & $0.8243 \pm 0.0147$        & -\\
    $r_2$                   & $0.0219 \pm 0.0005$                 & -\\
    $\rho_\star$ (\gcc)     & $6.42 \pm 0.63$                    & \\ 
    $K$ (m\,s$^{-1}$)       & $1.05 \pm 0.10$                     & $0.94 \pm 0.11$\\
    \hline
    \textbf{Derived parameters} &                     & \\
    \hline
    $a/R_\star$               & $11.23 \pm 0.37$ & - \\
    $R_p$ ($R_\oplus$)          & $1.130 \pm 0.043$   & - \\
    $b$                         & $0.737 \pm 0.022$   & - \\
    $I$ (degrees)               & $86.245 \pm 0.237$  & - \\
    $T_\mathrm{dur}$ (h)               & $0.988 \pm 0.013$   & - \\ 
    $T_\mathrm{eq}$ (K)                & $722 \pm 35$   & $533 \pm 26$ \\ 
    $M$sin$i$ ($M_\oplus$)      & $1.256 \pm 0.128$   & $1.527 \pm 0.181$ \\ 
    $M$ ($M_\oplus$)            & $1.258 \pm 0.128$   & - \\ 
    \hline
    \end{tabular}
    \caption{The planet parameters from the \texttt{juliet} fit to both the detrended \tess data and the MAROON-X RV data.}
    \label{tab:fit_joint_planet}
\end{table*}

\begin{table*}[]
    \centering
    \begin{tabular}{|l|c|c|c|c|c|c|}
    \hline
    \textbf{Transit parameters} &  & & & & & \\
    \hline
    Instrument     & $\mu$ (ppm)  & $\sigma$ (ppm)            & $q_{1}$                & $q_{2}$                & $u_{1}$         & $u_{2}$ \\
    \hline
    \tess$_{120s}$     & $-2 \pm 22$  & $15^{+21}_{-10}$   & $0.30^{+0.24}_{-0.17}$ & $0.34^{+0.40}_{-0.24}$ & $0.33 \pm 0.29$ & $0.15 \pm 0.32$ \\
    \tess$_{20s}$      & $10 \pm 18$  & $863^{+26}_{-26}$  & $0.32^{+0.17}_{-0.15}$ & $0.52^{+0.30}_{-0.30}$ & $0.54 \pm 0.30$ & $-0.03 \pm 0.30$ \\
Sinistro$_\mathrm{Jul}$& $26 \pm 105$ & $1140^{+77}_{-95}$ & $0.46^{+0.35}_{-0.30}$ & $0.60^{+0.28}_{-0.36}$ & $0.69\pm 0.46$  & $-0.09 \pm 0.39$ \\
Sinistro$_\mathrm{Aug}$& $2 \pm 81$   & $830^{+73}_{-89}$  & $0.73^{+0.17}_{-0.29}$ & $0.68^{+0.23}_{-0.31}$ & $1.09 \pm 0.45$ & $-0.28 \pm 0.40$ \\
    MuSCAT3$_g$        & $-3 \pm 61$  & $557^{+79}_{-73}$  & $0.54^{+0.31}_{-0.34}$ & $0.58^{+0.25}_{-0.35}$ & $0.75 \pm 0.45$ & $-0.09 \pm 0.38$ \\
    MuSCAT3$_r$        & $30 \pm 68$  & $1725^{+51}_{-60}$ & $0.49^{+0.31}_{-0.27}$ & $0.41^{+0.37}_{-0.27}$ & $0.48 \pm 0.42$ & $0.10 \pm 0.39$ \\
    MuSCAT3$_i$        & $24 \pm 74$  & $2243^{+61}_{-62}$ & $0.40^{+0.38}_{-0.27}$ & $0.52^{+0.32}_{-0.36}$ & $0.55 \pm 0.42$ & $-0.02 \pm 0.38$ \\
    MuSCAT3$_z$        & $27 \pm 58$  & $2098^{+47}_{-42}$ & $0.40^{+0.34}_{-0.27}$ & $0.43^{+0.36}_{-0.29}$ & $0.46 \pm 0.41$ & $0.07 \pm 0.37$ \\
    \hline
    \textbf{RV parameters} &  & & & & &\\
    \hline    
    Parameter & Red & Blue & & & & \\
    \hline
    $\mu_{\mathrm{Apr 21}}$ (m\,s$^{-1}$)                               & $-7.12 \pm 0.80$ & $-7.36 \pm 0.99$ & & & &  \\
    $\mu_{\mathrm{May 21}} - \mu_{\mathrm{Apr 21, R}}$(m\,s$^{-1}$)     & $2.81 \pm 0.75$  & $2.71 \pm 0.87$ & & & &  \\
    $\mu_{\mathrm{Aug 21}}$(m\,s$^{-1}$)                                & $-4.99 \pm 0.79$ & $-2.81 \pm 1.08$ & & & &  \\
    $\mu_{\mathrm{Nov 21}} - \mu_{\mathrm{Aug 21, R}}$(m\,s$^{-1}$)     & $2.44 \pm 0.77$  & $1.52 \pm 0.84$ & & & &  \\
    $\mu_{\mathrm{Mar 22}}$(m\,s$^{-1}$)                                & $1.45 \pm 0.57$  & $1.17 \pm 0.71$ & & & &  \\
    $\mu_{\mathrm{May 22}} - \mu_{\mathrm{Mar 22, R}}$(m\,s$^{-1}$)     & $1.64 \pm 0.77$  & $-0.87 \pm 0.92$ & & & &  \\
    $\mu_{\mathrm{Jul/Aug 22} - \mu_{\mathrm{Mar 22, R}}}$(m\,s$^{-1}$) & $3.31 \pm 0.69$  & $0.67 \pm 0.87$ & & & &  \\
    $\mu_{\mathrm{Jul 23}}$(m\,s$^{-1}$)                                & $6.84 \pm 0.79$  & $4.70 \pm 1.06$ & & & &  \\
    $\sigma_{\mathrm{Apr 21}}$(m\,s$^{-1}$)                             & $0.79^{+0.80}_{-0.72}$ & $0.03^{+0.34}_{-0.02}$ & & & &  \\     
    $\sigma_{\mathrm{May 21}}$(m\,s$^{-1}$)                             & $0.46^{+0.13}_{-0.22}$ & $0.51^{+0.45}_{-0.50}$ & & & &  \\         
    $\sigma_{\mathrm{Aug 21}}$(m\,s$^{-1}$)                             & $0.03^{+0.31}_{-0.03}$ & $0.20^{+1.13}_{-0.19}$ & & & &  \\         
    $\sigma_{\mathrm{Nov 21}}$(m\,s$^{-1}$)                             & $0.02^{+0.17}_{-0.02}$ & $0.04^{+0.26}_{-0.04}$ & & & &  \\         
    $\sigma_{\mathrm{Mar 22}}$(m\,s$^{-1}$)                             & $0.70^{+0.39}_{-0.64}$ & $0.05^{+0.30}_{-0.05}$ & & & &  \\         
    $\sigma_{\mathrm{May 22}}$(m\,s$^{-1}$)                             & $0.03^{+0.26}_{-0.03}$ & $0.37^{+1.13}_{-0.36}$ & & & &  \\         
    $\sigma_{\mathrm{Jul/Aug 22}}$(m\,s$^{-1}$)                         & $0.01^{+0.18}_{-0.01}$ & $0.06^{+0.49}_{-0.05}$ & & & &  \\  
    $\sigma_{\mathrm{Jul 23}}$(m\,s$^{-1}$)                             & $0.05^{+0.22}_{-0.05}$ & $0.02^{+0.12}_{-0.01}$ & & & &  \\         
    \hline
    \textbf{GP parameters} &  \\
    \hline 
    $\sigma_\mathrm{GP, red}$  & $1.40 \pm 0.22 $ & & & & & \\
    $\sigma_\mathrm{GP, blue}$ & $1.85 \pm 0.30 $ & & & & & \\
    $\alpha_\mathrm{GP}$       & $0.000025^{+0.000196}_{-0.000022}$& & & & & \\
    $\Gamma_\mathrm{GP}$       & $8.70 ^{+0.89}_{1.40}$& & & & & \\
    $P_\mathrm{rot}$           & $40.1^{+1.7}_{-1.5}$ & & & & & \\
    \hline
    \end{tabular}
    \caption{The instrument parameters from the \texttt{juliet} fit to both the detrended \tess data and the MAROON-X RV data.}
    \label{tab:fit_joint_instrument}
\end{table*}

\section{Discussion}
\label{sec:discussion}

\subsection{Dynamics}
\label{ssec:stability}

We must check to see if the reported planet system is stable before claiming a detection, especially given the fact that there is some ambiguity in the model selection.  An analysis that shows that the system is dynamically stable on long timescales would support the two-planet model selected in Section~\ref{ssec:rv_model}.  We can do so quickly with an analytical method inspired by \cite{Lissauer2011}, which makes use of the mutual Hill sphere radius:

\begin{equation}
    R_{H_{i, o}} = \bigg(\frac{M_i+ M_o}{3 M_\star}\bigg)^{1/3} \frac{a_i + a_o}{2} \
\end{equation}

where $i$ and $o$ are adjacent planet indices, with $i$ being an inner planet and $o$ being an outer planet.  The dynamical orbital separation $\Delta$ is defined as 

\begin{equation}
    \Delta = \frac{a_o-a_i}{R_{H_{i, o}}}
\end{equation}

If we assume that the planets are roughly coplanar and that sin$i \, \approx \, 1$, $\Delta_{cb}$ is approximately equal to 33.  As shown in \cite{Gladman1993}, a pair of small planets on circular orbits with $\Delta\,>\,2\sqrt{3}$ are stable against close approaches for very long periods of time.  This indicates that this system would be stable if there are no other planets in the system.  should be stable.  


If a planet is present at 3.18d or 3.39d (and thus in three-body resonance with the other two planets) $\Delta\,>9$ even if the 3d planet had a mass of $\geq\,10\,M_{\oplus}$.  If such a planet was present, it would also satisfy the stability boundary for three-planet systems discussed in \cite{Lissauer2011}, which would be fulfilled if 

\begin{equation}
    \Delta_{cb} + \Delta_{dc}\,>\,18.
\end{equation}

If this planet was present, the planets would be relatively evenly-spaced, aligning with the ``peas-in-a-pod" pattern of orbital spacing noted in the \kepler transiting planet sample by \cite{Weiss2018}.  With period ratios 1.5\,---\,1.7, the system would fall on the boundary between correlated and uncorrelated period ratios described in \cite{Jiang2020}, which concluded that evenly spaced period ratios are likely the result of mean motion resonances (MMRs).  If the TOI-1450A planetary system is the result of disk-driven inward planet migration, we may expect to see the presence of this additional planet.

Regarding the secondary star, we note that \cite{Ballantyne21} states that, if the binary stars TOI-1450A and TOI-1450B have a circular orbit, any planet orbiting the primary with $a\,\leq\,28.6$\,AU is stable with respect to the binary system.  As all of the planets discussed here have a measured $a$ well within this limit, it's clear that this system should be stable with regards to the companion star.

\subsection{Transiting Planet Composition}

With our joint fit, we find that the transiting planet has a mass of $M_p\, = \,1.26\,\pm\,0.13 M_\oplus$ and a radius of $R_p\,=\,1.13\,\pm\,0.04 R_\oplus$.  Given our knowledge of the mass and radius of planet, we can estimate its internal composition using the \texttt{ManipulatePlanet} tool \citep{Zeng13, Zeng16}.  \texttt{ManipulatePlanet} solves for the planet's structure based on an extrapolation based on Earth's seismic density profile.  It estimates the planet's iron, water, and rock contents, though these values are somewhat degenerate with the planet's central pressure, which is an unknown value.  We note that this code assumes an atmosphere is not present, which means that it may underestimate the iron content of the planet.

\begin{figure}
    \centering
    \includegraphics[width=0.9\linewidth]{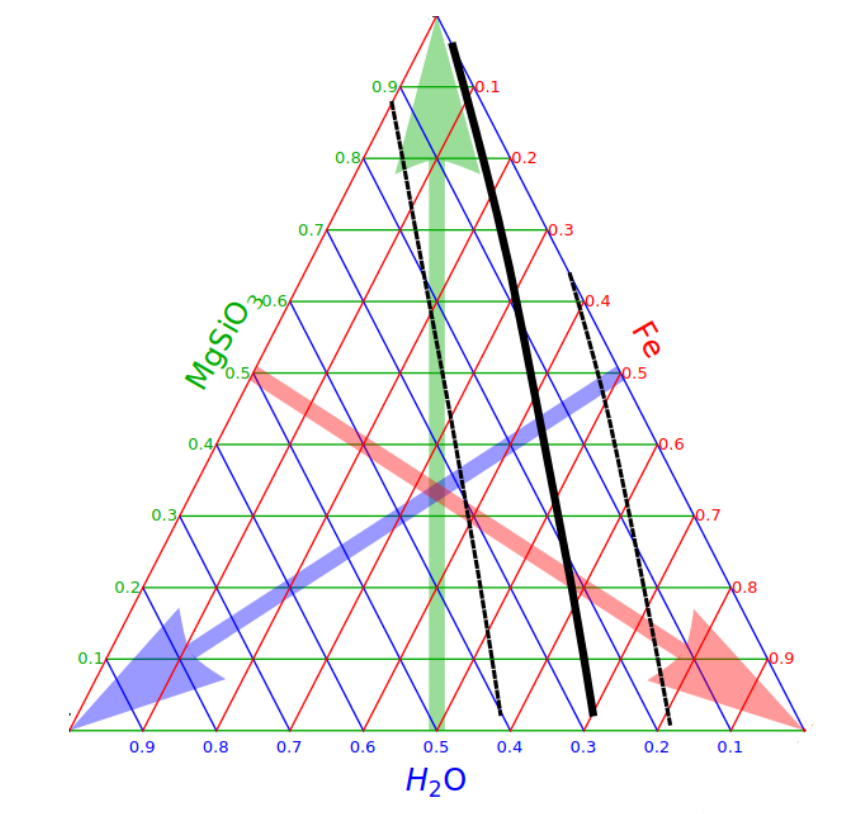}
    \caption{Ternary diagram of the composition of TOI-1450Ab generated using \texttt{ManipulatePlanet} \citep{Zeng13, Zeng16}.  The solid black line indicates the planet composition for a variety of different central pressure values, while the dotted black lines indicate the range of values given the mass and radius errors.}
    \label{fig:manipulate_planet}
\end{figure}

The resulting ternary diagram from \texttt{ManipulatePlanet} is shown in Figure~\ref{fig:manipulate_planet}. It appears that this planet may have a core mass fraction that is lower than that of the Earth \citep[$32.5\,\pm\,0.3\%$, from][]{Wang2018}, though mass and radius measurements are not precise enough to allow for a better composition estimate.  More precision on the mass and radius measurements will be necessary to prove whether or not this planet is less dense than the Earth.  TOI-1450Ab's low density could hint at the presence of an atmosphere, but could also just be a sign of a small core or an enhanced bulk volatile composition, which is possible if the planet formed at larger orbital separations and migrated inward \citep[see, e.g., the results of the simulations from][]{Burn2021}.  Even if it does possess an atmosphere, however, its high equilibrium temperature rules out the possibility of the planet being habitable, even when adopting the most optimistic HZ limits from \cite{Kopparapu16}.

\begin{figure}
    \centering
    \includegraphics[width=0.9\linewidth]{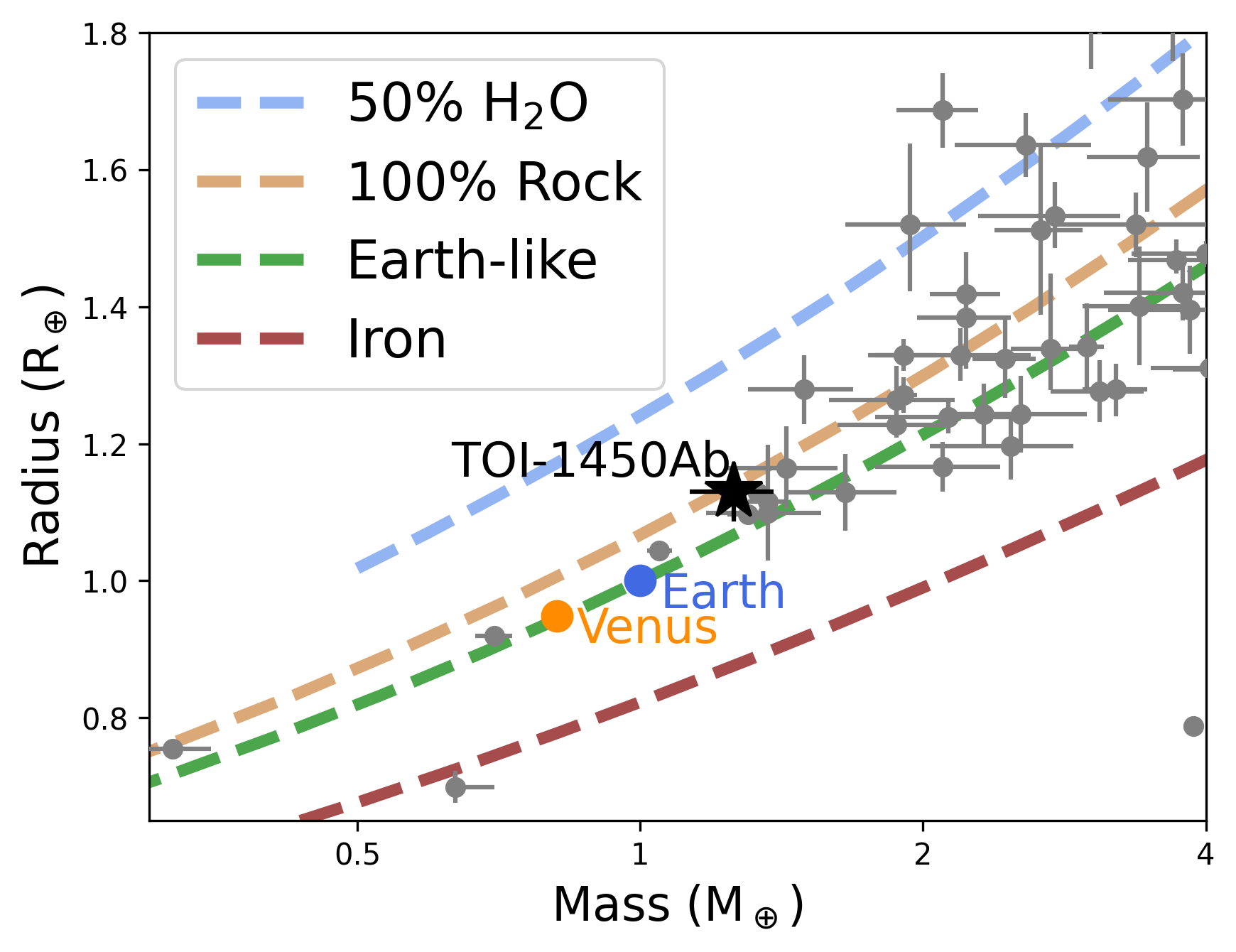}
    \caption{The mass and radius of TOI-1450Ab (black star) compared to both interior composition models from \cite{Zeng19} (colored dashed lines) and other exoplanets with mass and radius errors $<\,20\%$ from TEPCat (gray points).  Earth and Venus are also included for reference.}
    \label{fig:mass_radius}
\end{figure}

Figure~\ref{fig:mass_radius} shows where TOI-1450Ab sits on a mass-radius plot compared to other exoplanets downloaded from TEPCat \citep{Southworth11} as of May 2, 2024.  Composition models from \cite{Zeng19} are overplotted for comparison.  TOI-1450Ab appears to have a lower density than what we would expect from an Earthlike planet, indicating that it either lacks a core or has some meaningful volatile component.  If we adopt the stellar paramters from SED fitting, the planet has an even lower density.  However, a higher mass precision (and better stellar parameter characterization) will be necessary to make any confident conclusions about the planet's composition.  We also note that TOI-1450Ab is one of the lowest-mass planets with a sub-20\% mass error, with the only smaller (and more precise) planet masses being the TTV-derived masses of the TRAPPIST-1 system and the RV-derived mass of the ultra-short-period planet GJ 367b \citep{Goffo2023}.  TOI-1450Ab is the second lowest-mass planet with a precise RV mass.

We can only estimate the minimum mass of TOI-1450Ac.  If it is coplanar with TOI-1450Ab, with $i\,\approx \,86^o$, it has a mass of about 1.5 Earth masses, making it only slightly larger than the Earth.  It could be very similar in composition to TOI-1450Ab, though it is impossible to estimate its composition without a radius measurement.

\subsection{Suitability for Atmospheric Characterization}

Our measured mass precision for TOI-1450Ab is approximately 10\%.  As the measured error is less than 20\%, the planet's mass precision is now at the level that the planet is amenable to high-precision atmospheric characterization with \jwst \citep{Batalha19}.  However, it is first important to determine whether or not the planet is suitable for such a measurement.  \cite{Kempton18} created two metrics for the purpose of identifying transiting planets most suitable for atmospheric characterization.  These are primarily useful for prioritizing planets for follow-up studies with \jwst.  

The transmission spectroscopy metric (TSM) gives a description of how suitable the planet is for transmission spectroscopy.  Assuming a low albedo, TOI-1450Ab has a TSM of $14.3 \,\pm \,2.1$.  This value is slightly (but not significantly) higher if we assume the SED-derived stellar characteristics, with a TSM of $14.5\,\pm\,2.4$.  Given the planet's small radius, it is above the $TSM\,=\,12$ threshold described by \cite{Kempton18} for high-quality atmospheric characterization targets, though only at the $\approx\,1\sigma$ level.  As TOI-1450Ab has a very similar mass and radius to that of the Earth, it at least superficially appears that this target should be prioritized for follow-up atmospheric characterization efforts.  However, given its low mass and short orbital period, it is possible that any substantial atmosphere that the planet may have had has already been lost.  

We can estimate the photoevaporative mass loss rate of the planet's atmosphere given the following equation from \cite{Sanz-Forcada2011}:

\begin{equation}
\dot{M} = \frac{\pi R_p^3 F_\mathrm{XUV}}{\mathrm{G} K M_p}
\end{equation}

where $F_\mathrm{XUV}$ is the XUV flux of the host star at the planet's orbit, G is the gravitational constant, and $K$ is a term that describes the potential energy reduction due to tidal forces, using the formula from \cite{Erkaev2007}.  

While there are no direct $F_\mathrm{XUV}$ measurements of the host star, \cite{Freund2022} identified that a point X-ray source from the Second ROSAT all-sky survey \citep[2RXS][]{Boller2016} seemed to roughly correspond with the known right ascension and declination of TOI-1450A.  However, the X-ray flux of this source ($2.56*10^{-14}$\,erg/s/cm$^2$) is close to the detection limit of the instrument and the resolution of the instrument is low enough that this source likely also includes the light from TOI-1450B, so we can consider this X-ray flux to be an upper limit.  Given the star's distance from \cite{GaiaEDR3}, this corresponds to an X-ray luminosity $L_\mathrm{X}$ of $1.54*10^{27}$\,erg/s.  As there are no available EUV flux measurements of the host star, we used the relation from \cite{Sanz-Forcada2011} to estimate the EUV luminosity, finding $L_\mathrm{EUV} \,\approx\, 1.5*10^{28}$\,erg/s, though we note that this relation has a high degree of scatter.  This results in a $F_\mathrm{XUV}$ at the planet's orbit of about $9.7*10^3$ erg/s/cm$^2$.  Plugging these numbers into the above equation, we find that $\dot{M}\,\approx\,0.12$\,\MEarth/Gyr.  This mass loss rate seems to be high enough that it would indicate that the planet could lose any substantial atmosphere over the course of its life, especially given that the host star likely had a higher XUV flux earlier in its lifetime.  However, we note that the X-ray flux is a likely upper limit and the method used to estimate the EUV flux is imprecise, so it is certainly possible that the actual mass loss rate is much slower.  The current data are insufficient to conclude whether or not TOI-1450Ab has lost its atmosphere. 

We can evaluate the likelihood of TOI-1450Ab of having an atmosphere by comparing it to similar planets with \jwst measurements.  TRAPPIST-1b, which superficially resembles TOI-1450Ab in that it is a roughly Earth-sized M dwarf planet at a sub-3d orbital period \citep[see, e.g.,][]{Agol2021}, was recently observed by \jwst and found to have little to no atmosphere \citep{Greene2023}.  As TOI-1450Ab has an even higher equilibrium temperature than TRAPPIST-1b, it seems possible that it was able to undergo more extreme atmospheric loss.  However, TRAPPIST-1 has a much later spectral type than TOI-1450A, which could be influential on the planet's atmosphere due to the star's enhanced activity levels and longer activity lifetime \citep{Hawley00}.  TOI-1450Ab is a challenging target with MIRI/LRS as envisioned by \cite{Kempton18}.  However, \cite{Greene2023} and \cite{Zieba2023} have been able to successfully detect the thermal emission of TRAPPIST-1b and TRAPPIST-1c (which have lower emission spectroscopy metrics) at 15 microns, demonstrating that these challenging targets can still be characterized with the at longer wavelengths.

\section{Summary and Conclusions}
\label{sec:conclusions}

TOI-1450 is a binary star system with a mid-M primary and a late-M secondary component.  The two stars are too close to separate on an individual \tess pixel, complicating efforts at follow-up given the secondary's strong rotation signal.  However, by performing a follow-up RV survey of TOI-1450A with MAROON-X, we were able to confirm the presence of a 2.044d planet signal that transits the primary star at the correct period and phase, and found that it is consistent with a $1.13\,\pm\,0.04$\,\REarth planet with a mass of $1.26\,\pm\,0.13$\,\MEarth.  This planet has a density consistent with Earth \citep[given the models from][]{Zeng19}, but could also possess some volatiles or even an atmospheric envelope given its slightly sub-Earth density.  Given its mass and radius, TOI-1450Ab is above the cutoff for suitability for atmospheric transmission spectrum characterization with \jwst, which may provide us with an interesting opportunity to study the atmosphere of a hot, low-density M dwarf planet.  However, given the planet's short orbital period, it is unclear whether or not TOI-1450Ab would be able to retain an atmosphere over long periods of time, necessitating further modeling efforts.  We also discovered the presence of an additional non-transiting planet in the system with a 5.07\,day period and and a $M_p$sin$i_p$ similar to TOI-1450Ab.

In addition, while the photometric data contains a sinusoidal 2d signal that is likely a rotation signal, we were able to use a combination of spectroscopic activity indicators and chromatic RV signals to support the hypothesis that TOI-1450B is the rapid rotator in the system.  Additionally, spectroscopic activity indicators and undetrended \tess photometry support a roughly 40d rotation period for TOI 1450A- a long enough rotation period that the star's rotation is unlikely to strongly influence our results.

\vskip 5.8mm plus 1mm minus 1mm
\vskip1sp

This material is based upon work supported by the National Science Foundation Graduate Research Fellowship under Grant No. DGE 1746045. GS acknowledges support provided by NASA through the NASA Hubble Fellowship grant HST-HF2-51519.001-A awarded by the Space Telescope Science Institute, which is operated by the Association of Universities for Research in Astronomy, Inc., for NASA, under contract NAS5-26555.  This work is partly supported by JSPS KAKENHI Grant Number JP18H05439 and JST CREST Grant Number JPMJCR1761.  This research has also made use of NASA's Astrophysics Data System Bibliographic Services.

The University of Chicago group acknowledges funding for the MAROON-X project from the David and Lucile Packard Foundation, the Heising-Simons Foundation, the Gordon and Betty Moore Foundation, the Gemini Observatory, the NSF (award number 2108465), and NASA (grant number 80NSSC22K0117). The Gemini observations are associated with programs GN-21A-Q-120, GN-21B-Q-216, GN-22A-Q-120, GN-22A-Q-218, GN-22B-Q-214, and GN-23A-Q-120 (PI: Bean).


Resources supporting this work were provided by the NASA High-End Computing Capability (HECC) Program through the NASA Advanced Supercomputing (NAS) Division at Ames Research Center for the production of the SPOC data products.


This work makes use of observations from the LCOGT network. Part of the LCOGT telescope time was granted by NOIRLab through the Mid-Scale Innovations Program (MSIP). MSIP is funded by NSF.


This research has made use of the Exoplanet Follow-up Observation Program (ExoFOP; DOI: 10.26134/ExoFOP5) website, which is operated by the California Institute of Technology, under contract with the National Aeronautics and Space Administration under the Exoplanet Exploration Program.


Funding for the TESS mission is provided by NASA's Science Mission Directorate. KAC and CNW acknowledge support from the TESS mission via subaward s3449 from MIT. We acknowledge the use of public TESS data from pipelines at the TESS Science Office and at the TESS Science Processing Operations Center.


This paper is based on observations made with the MuSCAT3 instrument, developed by the Astrobiology Center and under financial supports by JSPS KAKENHI (JP18H05439) and JST PRESTO (JPMJPR1775), at Faulkes Telescope North on Maui, HI, operated by the Las Cumbres Observatory.


Some of the data presented in this paper were obtained from the Mikulski Archive for Space Telescopes (MAST) at the Space Telescope Science Institute. The specific observations analyzed can be accessed via \cite{TESS_Curves}.


\software{AstroImageJ \citep{Collins:2017}, Astropy \citep{Astropy1, Astropy2, Astropy3}, astroquery \citep{astroquery}, batman \citep{batman}, dynesty \citep{Speagle2020}, emcee \citep{emcee}, exoplanet \citep{exoplanet:exoplanet}, juliet \citep{Espinoza19}, lightkurve \citep{lightkurve}, Numpy \citep{numpy}, PyAstronomy \citep{PyAstronomy}, PyMC3 \citep{exoplanet:pymc3}, Scipy \citep{2020SciPy-NMeth}, spectres \citep{Carnall17}, TAPIR\citep{Jensen:2013}, theano \citep{exoplanet:theano}, tpfplotter \citep{Aller20}}

\facility {Exoplanet Archive, LCOGT}

\bibliography{manuscript}
\end{document}